%% file: ms.tex
\documentclass[twocolumn]{aastex701}

\usepackage[utf8]{inputenc}
\usepackage{amsmath,amssymb}
\usepackage{tikz}
\usetikzlibrary{arrows.meta, positioning, fit}

\newcommand{\mjup}{\ensuremath{\mathrm{M_{jup}}}}
\newcommand{\msun}{\ensuremath{\mathrm{M}_\odot}}

\newcommand{\feh}{\ensuremath{[\mathrm{M/H}]}}
\newcommand{\paperI}{Paper I}
\newcommand{\paperII}{Paper II}

\renewcommand{\P}{\mathbb{P}}
\newcommand{\E}{{\mathbb{E}}}
\newcommand{\I}{{\mathbb{I}}}
\newcommand{\mciter}{^{(i)}}
\newcommand{\Nmax}{N_\mathrm{max}}
\newcommand{\psiint}{\psi^\mathrm{int}}
\newcommand{\piint}{\pi^\mathrm{int}}

\newcommand{\nrc}{NRC Herzberg Astronomy and Astrophysics,
5071 West Saanich Road,
Victoria, BC, V9E 2E7, Canada}
\newcommand{\ucsd}{Department of Astronomy \& Astrophysics,  University of California, San Diego, La Jolla, CA 92093, USA}
\newcommand{\nmsu}{Department of Astronomy, New Mexico State University, 1320 Frenger Mall, Las Cruces, NM 88003, USA}
\newcommand{\uvic}{Department of Physics and Astronomy, University of Victoria,
3800 Finnerty Road, Elliott Building,
Victoria, BC, V8P 5C2, Canada}
\newcommand{\mac}{Department of Physics \& Astronomy, McMaster University, 1280 Main Street West, Hamilton, ON, L8S 4L8, Canada}
\newcommand{\caltech}{Department of Astronomy, California Institute of Technology, Pasadena, CA 91125, USA}

\newcommand{\ucsc}{Department of Astronomy \& Astrophysics, University of California, Santa Cruz, CA 95064, USA}
\newcommand{\ucsb}{Department of Physics, University of California, Santa Barbara, CA 93106, USA}

\shorttitle{The brown dwarf desert is a mass-ratio desert}
\shortauthors{Thompson et al.}
\submitjournal{ApJL}

\begin{document}

\title{The Brown Dwarf Desert is a Mass-Ratio Desert: Astrometric Demographics of 63,254 Stars}

\author[0000-0001-5684-4593]{William Thompson}
\affiliation{\nrc}
\email[show]{william.thompson@nrc-cnrc.gc.ca}

\author[0000-0001-7402-8506]{Alexandre Bouchard-Côté}
\affiliation{Department of Statistics, University of British Columbia, Vancouver, BC V6T 1Z4, Canada}
\email{bouchard@stat.ubc.ca}

\author[0000-0001-9582-4261]{Dori Blakely}
\affiliation{\uvic}
\affiliation{\nrc}
\email{blakelyd@uvic.ca}

\author[0000-0003-1499-0191]{Trevor Campbell}
\affiliation{Department of Statistics, University of British Columbia, Vancouver, BC V6T 1Z4, Canada}
\email{trevor@stat.ubc.ca}

\author[0000-0003-3430-3889]{Jessica Speedie}
\altaffiliation{Heising-Simons Foundation 51 Pegasi b Fellow}
\affiliation{Department of Earth, Atmospheric, and Planetary Sciences, Massachusetts Institute of Technology, Cambridge, MA 02139, USA}
\email{jspeedie@mit.edu}

\author[0000-0002-4290-6826]{Judah Van Zandt}
\affiliation{\ucsb}
\email{judahvz@ucsb.edu}

\author[0000-0003-4557-414X]{Kyle Franson}
\altaffiliation{NHFP Sagan Fellow}
\affiliation{\ucsc}
\email{kfranson@ucsc.edu}

\author[0000-0002-6618-1137]{Jerry W. Xuan}
\altaffiliation{Heising-Simons Foundation 51 Pegasi b Fellow}
\affiliation{\ucsb}
\email{jxuan@ucsb.edu}

\author[0000-0002-6773-459X]{Doug Johnstone}
\affiliation{\nrc}
\affiliation{\uvic}
\email{Douglas.Johnstone@nrc-cnrc.gc.ca}

\author[0000-0002-1845-2617]{Emma V. Turtelboom}
\affiliation{Department of Physics \& Astronomy, McMaster University, 1280 Main Street West, Hamilton, Ontario, Canada}
\email{turtlebe@mcmaster.ca}

\author[0000-0002-2696-2406]{Jingwen Zhang}
\affiliation{\ucsb}
\email{jwzhang@ucsb.edu}

\author[0000-0001-5173-2947]{Clarissa R. Do \'O}
\affiliation{\caltech}
\email{cdoo@caltech.edu}

\author[0000-0003-2649-2288]{Brendan P. Bowler}
\affiliation{\ucsb}
\email{bpbowler@ucsb.edu}

\author[0000-0003-2233-4821]{Jean-Baptiste Ruffio}
\affiliation{\ucsd}
\email{jruffio@ucsd.edu}

\author[0000-0001-6975-9056]{Eric Nielsen}
\affiliation{\nmsu}
\email{nielsen@nmsu.edu}

\author[0000-0001-5383-9393]{Ryan Cloutier}
\affiliation{\mac}
\email{ryan.cloutier@mcmaster.ca}

\author[0009-0007-6766-2040]{Andre Fogal}
\affiliation{\uvic}
\email{afogal@uvic.ca}

\author[0009-0009-2223-2404]{Kaitlyn Hessel}
\affiliation{\uvic}
\email{khessel@uvic.ca}

\author[0000-0002-4164-4182]{Christian Marois}
\affiliation{\nrc}
\affiliation{\uvic}
\email{Christian.Marois@nrc-cnrc.gc.ca}

\author[0009-0008-1229-3230]{Alexandra Rochon}
\affiliation{\mac}
\email{rochoa3@mcmaster.ca}

\author[0009-0008-9687-1877]{William Roberson}
\affiliation{\nmsu}
\email{wcr@nmsu.edu}

\author[0000-0002-9820-1884]{Simon Blouin}
\affiliation{\uvic}
\email{sblouin@uvic.ca}

\begin{abstract}
Brown dwarfs are rare as close companions to stars compared to planets and stellar companions, a paucity known as the brown dwarf desert.
Whether the desert sits at a fixed companion mass or a fixed companion-to-star mass-ratio $q$ has been difficult to test, because microlensing, radial velocities, and direct imaging have sampled comparatively narrow slices of stellar mass and separation.
Here we apply a novel joint-modelling method to Hipparcos and Gaia DR2/DR3 astrometry from 63,254 stars to complete this picture with a sample 1-2 orders of magnitude larger than those of previous wide-orbit demographic studies. For the first time, we present a single survey that measures occurrence rates of both planetary and stellar companions on 3--94\,yr orbits ($\sim$2--20\,au) across host masses from M dwarfs to B stars. We find that the desert is a mass-ratio feature: companions are strongly depleted at $q\approx2\%$ independent of stellar mass from M to B stars, and a desert at fixed companion mass is excluded. Furthermore, we find a positive trend in occurrence with metallicity below the mass-ratio desert, but a negative trend above.
The desert therefore separates two distinct populations, plausibly two formation channels, that each scale with stellar mass: the lightest stellar-like companions of K stars ($\sim13$ \mjup{}) are less massive than the heaviest planet-like companions of B stars ($\sim27$ \mjup{}), which continue well beyond the deuterium-burning limit. Mass-ratio, not mass, is therefore the physically meaningful boundary between planets and their stellar-like counterparts.

\end{abstract}


\section{Introduction}\label{sec:intro}

Brown dwarfs, with masses between the heaviest giant planets and the lightest stars, are rarely found as companions close to other stars. They are found around fewer than 1\% of Sun-like stars within $\sim5$ au by radial velocity surveys \citep{marcy_butler_2000,grether_2006,sahlmann_2011,cui_2026}, far fewer than either giant planets or stellar companions. This ``desert'' may be the gap between two formation channels: bottom up formation like core accretion and top down formation through fragmentation  \citep{ma_ge_2014,schlaufman_2018,chabrier_2014}. It has been shown that companion trends with metallicity reverse on opposite sides of the desert, strengthening this view \citep{schlaufman_2018,giacalone_2025}.

Earlier work placed the desert minimum near $\sim$30\,\mjup{} for roughly solar-mass hosts \citep{grether_2006}. Reviews of imaging and radial-velocity results have suggested that this minimum could scale with the host mass, so that the mass ratio may be the key discriminator \citep{currie_2023_ppvii}.
If both formation channels scale with the host mass, the desert may sit at a fixed companion-to-host mass-ratio $q$ rather than a fixed mass. There are hints at this from both ends of the main sequence: microlensing reveals a desert at $0.02 \lesssim q \lesssim 0.05$ around 0.05--0.8\,\msun{} hosts at 1--5\,au \citep{zhang_2025_desert}, and imaging finds the same $q$ range empty around 1.75--4.5\,\msun{} stars at 20--1000\,au \citep{duchene_2023}---though a rising occurrence below this $q$ is not yet established. Each method reaches a different separation around quite different host-masses, and no method has yet spanned low to high mass-ratio companions across a wide range of hosts. Whether the desert, with associated rising companion rates above and below, extends beyond the reach of RV and is indeed a mass-ratio feature across the main sequence, has not been tested.

We address this gap using astrometry. As companions orbit their stars, they pull the photocenter in an ellipse around their common center of mass, which can be detected by observatories such as Hipparcos \citep{hipparcos_1} and Gaia \citep{gaia_mission}. Opposite to RV, astrometry is more sensitive to planets on wider orbits, with sensitivity only dropping once the  orbital period exceeds the temporal baseline of the data ($\approx25\,$yr). The sensitivity of astrometry to companions is also fairly uniform across spectral types, as it does not require spectral lines to measure a Doppler shift and is not strongly impacted by stellar activity. Instead, it is principally limited by the distance to the star and its overall brightness. So far astrometry has mainly been used to determine the dynamical masses of individual companions found by imaging and radial velocity \citep[e.g.][]{Snellen2018, currie_2023, thompson_eps_eri}, to refine the masses of RV-detected planets \citep{An2025, cui_2026,VanZandt_2026}, to measure the 3-D orbital architectures of planetary systems \citep[e.g.][]{Xuan2020,Zhang2025}, or in a few cases, to detect new planets \citep[e.g. Gaia-4b;][]{stefansson_2025_gaia4b}. The imminent release of Gaia DR4 epoch astrometry is set to change this, with thousands of planet detections expected from Gaia DR4 alone \citep{Lammers2026}. Ahead of this release, this Letter presents the first population-level planet demographics with astrometry as the primary constraint, using data from Gaia DR2, DR3, and Hipparcos as tabulated in the ``G23H'' catalog of \citet[][hereafter \paperII{}]{thompson_apjs_2026}.

Measuring occurrence rates from these data brings its own challenge. Despite the large sample size, most candidate companions have poorly constrained orbital periods, and few can be independently confirmed as planets (see \citet{doo_g23h} for some exceptions). Rather than set a detection threshold, we therefore propagate each system's uncertainty over the number of companions and their parameters using a hierarchical model of the measurements of every star, rather than of a catalog of detected companions \citep{hogg_2010,foreman-mackey_2014}. This infers the population that best explains all systems at once. The implicit-completeness approach is related to the posterior-based occurrence framework recently developed for radial velocities by \citet{faria_inferring_2025} and applied in \citet{parc_2026}.

Using this new methodology and a significantly larger sample of 63,254 stars---two orders of magnitude larger than previous demographics studies at wide separations---we show that the brown dwarf desert is a mass-ratio feature at 2\% of the host mass on 3--94\,yr orbits from stars with host masses spanning 0.16–8.1 \msun{} (99\% lying between 0.54–5.1 \msun{}), and that the population of giant planets continues smoothly across this range, even where these companions exceed the deuterium-burning limit.

\begin{figure*}[p]
\centering
\vspace{-1em}
\includegraphics[width=\textwidth]{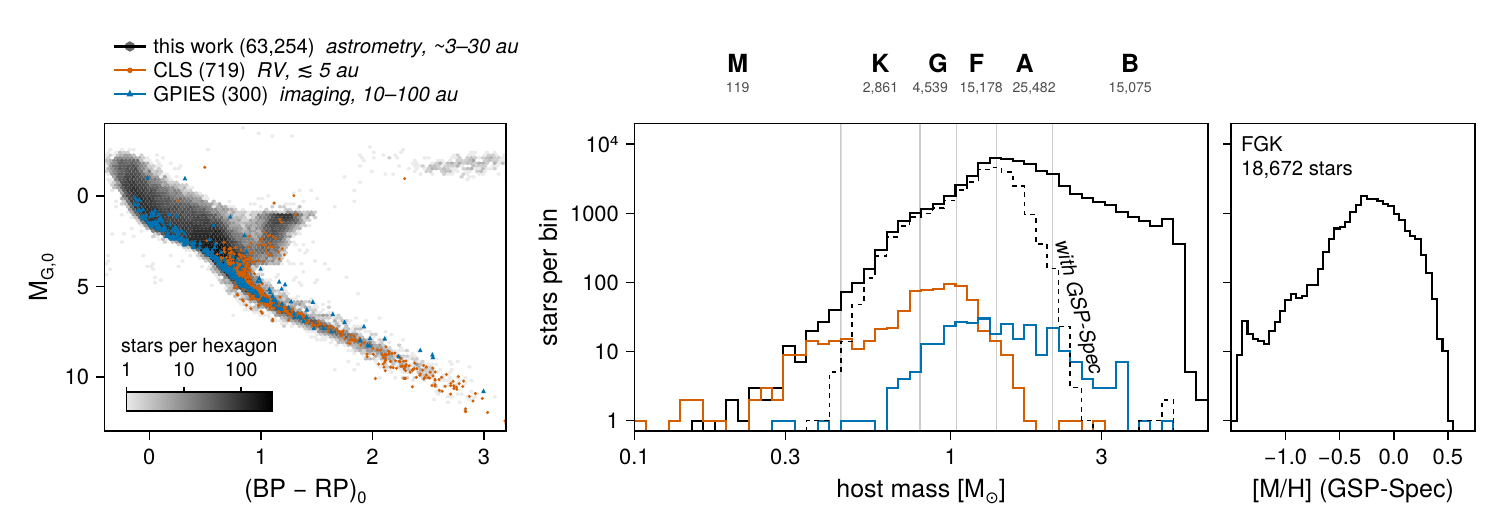}
\vspace{-1em}
\caption{The sample used in this work compared with two illustrative planet-demographics surveys from RV \citep[CLS, 719 stars;][]{Rosenthal2021, fulton_2021} and imaging \citep[GPIES, 300 stars;][]{nielsen_2019}. Astrometry from Gaia and Hipparcos can access a sample two orders of magnitude larger. \textbf{Left:} extinction-corrected Gaia color--magnitude diagram of our 63,254 pre core-He burning stars (gray hexagons, logarithmic counts), with the CLS and GPIES targets overplotted. \textbf{Center:} host-mass distributions. The dashed line shows the 28,001 of our stars with a usable GSP-Spec \feh{} (Section~\ref{sec:data}). Gray lines mark our host-mass bin edges at 0.45, 0.80, 1.04, 1.40 and 2.10\,\msun{}, and the letters and numbers above the panel give each bin's label and number of stars. \textbf{Right:} calibrated Gaia DR3 GSP-Spec \feh{} of the 18,672 FGK stars used in the metallicity analysis.}
\label{fig:sample}
\end{figure*}

\begin{figure*}[p]
\centering
\vspace{-0.8em}
\includegraphics[width=.8\textwidth]{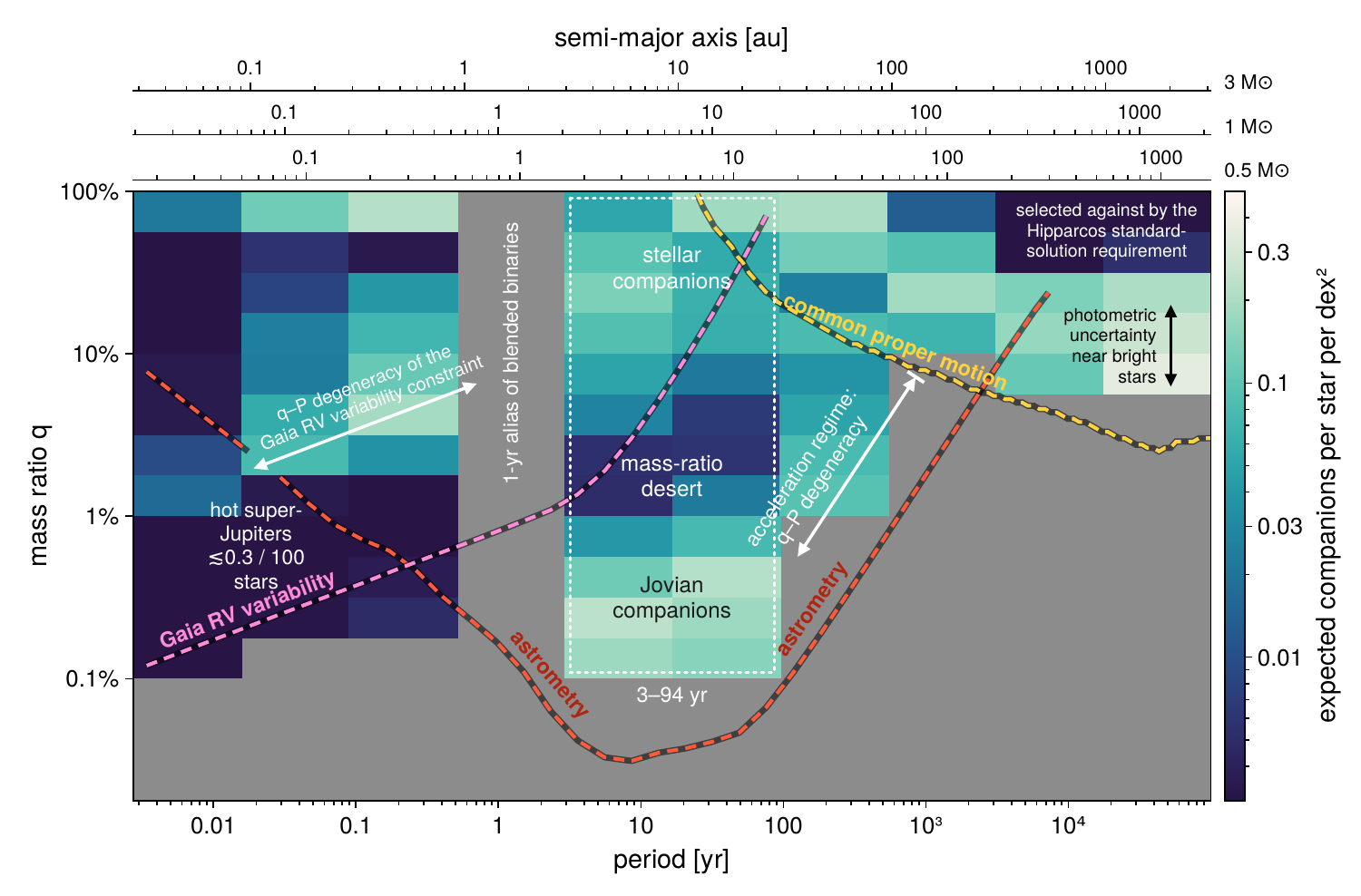}
\vspace{-1em}
\caption{The inferred population of companions around our full sample of 63,254 stars ranging from 0.16–8.1 \msun{} (99\%: 0.54–5.1): the posterior median of the expected number of companions per star, per dex of orbital period, per dex of mass ratio, from the hierarchical population model. A ``desert'' in occurrence at the center of the plot is seen to be a feature in mass-ratio across the full sample.  The masked cells have poor sensitivity (fewer than 300 stars are sensitive; see Section~\ref{sec:results}) or fall in the 1 yr column. A dotted line marks the 3--94\,yr window we consider in this work. The dashed lines show where different sources of information from Gaia and Hipparcos become informative  for the 300th most sensitive star in the sample: astrometry (red), Gaia DR3 RV variability (pink), and resolution of the companion as a separate DR3 source (yellow). White arrows mark the mass--period degeneracies of the RV-variability and acceleration constraints; the latter is broken where the companion would be resolved. The scales along the top give the semi-major axis corresponding to each period for host masses of 0.5, 1 and 3\,\msun{}.}
\label{fig:funnel}
\end{figure*}

\section{Data and Methods}\label{sec:data}

\paragraph{System posteriors.} We used the catalog from \citet{thompson_apjs_2026} (\paperII{}), which consists of posteriors fit to Gaia DR2, DR3, and Hipparcos data from the G23H catalog \citep[\paperI{};][]{Thompson_2026}.
The main constraint on companion orbits in the 2.9--94.4\,yr period range we study (the edges of two cells of the population grid described below) is astrometry from the Gaia DR2, DR3, and Hipparcos proper motion anomalies and from the Gaia astrometric excess noise (as introduced by \citealt{pmex1}, here using the calibration in \paperI{}, \S2.3). In addition to astrometry of the host star, the Gaia data also reveal a star's RV variability over the DR3 baseline (calibrated by \citet{paired}; see \paperII{}). \paperII{} also requires any wide stellar companion to match a co-moving Gaia source or to be too faint for Gaia ($G \gtrsim 22$), which breaks degeneracies in wide orbits (we do not consider stellar remnants). Finally, the mass of the host and any unresolved stellar companions is constrained based on the Gaia DR3 BP, RP, and G band photometry via the MIST isochrone grids \citep{mist_choi}, and marginalized over wide priors on the system's age and [Fe/H] (all components assumed to be coeval).

As discussed in \paperII{}, this combination of data is able to detect or rule out stellar companions from close-in spectral binaries to wide companions at 1000s of au, is complete for brown dwarfs between $\sim$2 and $\sim$80\,yr periods out to 100\,pc (at least 90\% can  exclude a 13\,\mjup{} companion), and is sufficiently sensitive to detect or rule out a 3\,\mjup{} giant planet around 58\% of stars in the sample at their most sensitive orbital periods (a 1.0 Jupiter-mass planet around 17\% of stars). For each star, the model marginalizes over 0 to 3 companions: the MCMC sampler moves between different numbers of companions at each draw in proportion to the prior on multiplicity and the evidence of the data.

\paragraph{Sample.} We begin with the ``recommended'' sample of 96,915 stars from \paperII{}, which are targets in Hipparcos, Gaia DR2, and DR3 with sufficiently common colors and photometry to calibrate the astrometric excess noise (\paperI{}, \S2.4).
This is a magnitude-limited ($3 \lesssim G \lesssim 12$) and volume-limited ($<1$\,kpc) sample. The bright end is set by Gaia, which is incomplete below $G \approx 7$ and misses most stars brighter than $G \approx 3$ \citep{gaia_dr2}; the faint end is set by Hipparcos, whose survey is complete to $V \approx 7.3$--9.0 but which also observed many fainter stars of particular interest \citep{hipparcos_1}. Requiring a five-parameter DR2 solution and a DR2--DR3 cross match, and excluding non-standard Hipparcos solutions, selects mildly against wide or partially resolved binaries \citep{lindegren_dr2_astrometry}.

\paragraph{Evolutionary Stage.} The photometric model used in \paperII{} assumes that no component has evolved past core helium burning. Here, we further only keep stars where more than half of their draws fall on the pre-main-sequence, main-sequence or subgiant/red-giant-branch phases of the MIST grid. This leaves 63,254 stars (65\% of the \paperII{} sample). The results are not sensitive to details of this cut on evolutionary stage.

\paragraph{Metallicity.} To study companion occurrence as a function of metallicity, we use Gaia DR3 GSP-Spec \feh{} \citep{recio_blanco_2023}, calibrated with the published $\log g$ polynomial. We keep stars with clean quality flags, a 16--84\% half-width below 0.2\,dex, and $3500 < T_\mathrm{eff} < 7500$\,K, leaving 28,001 stars with a usable \feh{} (29\% of the \paperII{} recommended sample). The metallicity analysis uses the 18,672 of these with host masses in our FGK bins (0.45--1.40\,\msun{}, after the reweighting below) that also pass the sensitivity selection described below. Because the \paperII{} \feh{} posteriors are unconstrained, we importance-reweight each star's draws by the likelihood of its GSP-Spec \feh{} (width $\sqrt{\sigma^2 + 0.1^2}$), which makes $M_\mathrm{pri}$ and $q$ consistent with the measured \feh{}. The GSP-Spec [M/H] scale is 0.13 dex metal-poor compared to APOGEE DR17 \citep{apogee_dr17} on the subset of stars in both catalogs. We report results on the GSP-Spec scale without correcting for this.

\paragraph{Host Mass.} The host masses for all stars in the sample are adopted directly from the posterior median of $M_{\mathrm{pri}}$ of the independent system-level posteriors of \paperII{}. For likely single stars, these masses agree with external estimates at the few-percent level (\paperII{}, \S3.3.3). If metallicity is incorporated from GSP-Spec, the shift in host mass compared to the Geneva--Copenhagen Survey \citep[GCS;][]{casagrande_2011} is just $-0.001$\,dex, or $+0.026$\,dex versus APOGEE StarHorse \citep{queiroz_2023}. We divide the sample into 6 bins in host mass: below 0.45 (``M''), 0.45–0.80 (``K''), 0.80–1.04 (``G''), 1.04–1.40 (``F''), 1.40–2.10 (``A''), and above 2.10 \msun{} (``B'').

\vspace{1em}
The full sample color--magnitude diagram, host-mass distribution, and metallicity distribution of this work is presented in Figure \ref{fig:sample}.

\paragraph{Sensitivity selection.} For the comparisons where the full sample is divided by host mass, we keep only stars that are sufficiently nearby to detect an approximately $q\leq0.01$ companion. This is to avoid the situation where a model fit to a smaller pool of stars ends up dominated by very distant stars with poor to no sensitivity. This leaves 119 M, 2,691 K, 4,168 G, 12,161 F, 14,873 A, and 2,828 B stars (36,840 total across all bins). The full-sample fit uses all 63,254 stars at once.

\paragraph{Population model.}
We fit a hierarchical Bayesian model whose population parameters are the distribution $\psi$ of the number of companions per star within the grid and the distribution $\pi$ of companions over a grid in log period ($\log P$) and log mass-ratio ($\log q$). We work in these coordinates since they have the most uniform sensitivity across host masses.  We fit $\pi$ as a grid of 10 equally spaced bins in $\log{P}$ between 1 d and $10^5$ yr, and 20 equally spaced bins in $\log{q}$ between $10^{-5}$  and 1.
The hierarchical model incorporates the \paperII{} system model of every star in the sample; the full model and sampler are described in Appendix~\ref{ap:methods}.

Fitting all systems jointly provides a ``super-resolution'' effect where inferences about the companions of each system are sharpened by what the rest of the sample reveals about the population, and signals of any strength---including those too weak to be detected independently---pool their information about the population.

Classical completeness corrections weight each detection by its inverse detection probability, a Horvitz--Thompson-type estimator, and hierarchical analyses typically plug in a separately estimated completeness \citep[e.g.,][]{foreman-mackey_2014}; we instead model each star's data jointly, following the model-based approach to survey inference \citep{little_2004}. Our model extends that of \citet{faria_inferring_2025}: as in that work, propagating each system's uncertainty over its companions accounts for incompleteness without injection--recovery tests (Appendix~\ref{ap:surveys}), provided the system model describes each star's sensitivity and the sample is not selected on the presence of companions (see Section~\ref{sec:data} and Figure \ref{fig:sample} for our sample). In other words, a posterior must be provided for every system, regardless of whether a companion was ``detected''.

To sample from the joint hierarchical posterior, we use a fast MCMC algorithm developed in the hierarchical Bayesian inference literature \citep{lunn_2013}.
Each system's update approximates an independence Metropolis--Hastings step by proposing one of its \paperII{} posterior draws; the likelihood of each system's data cancels in the acceptance ratio, so each update costs only a ratio of prior densities. The same ratio of population prior to \paperII{} prior underlies the importance-sampling approach of \citet{hogg_2010}; here it enters an MCMC acceptance step rather than a reweighting, so the population and the individual systems are sampled jointly (see Appendix~\ref{sec:imh-sampler}).

\begin{figure}[t]
\centering
\includegraphics[width=\columnwidth]{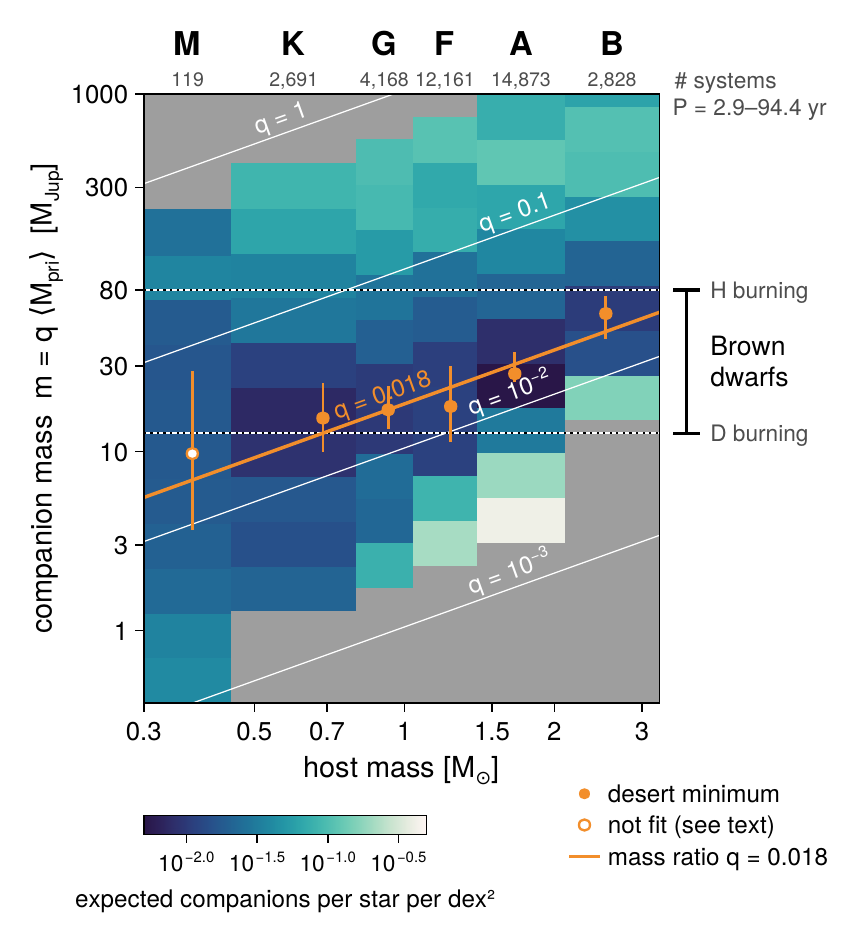}
\caption{Companion rates at 3--94\,yr as a function of host mass. We fit separate population models as functions of ($\log{P}, \log{q})$ to each host mass range. We convert the $\log{q}$ to mass $m$ via the mean host mass $<M_{\rm{pri}}>$ of each column. We see that the brown dwarf desert scales with the host mass, with the deepest point near 2\%. The stellar-companion side of the desert around low mass stars in fact overlaps with the planetary-companion side of the desert around high mass stars, showing that a valley in companion mass would not distinguish the two populations across the full range of host masses. For these fits, each column only uses stars that are potentially sensitive to $q\leq0.01$ companions (Section~\ref{sec:data}).
The open point (M) is not used for the fixed-mass-ratio line.}
\label{fig:desertmap}
\end{figure}

\begin{figure}[t]
\centering
\includegraphics[width=\columnwidth]{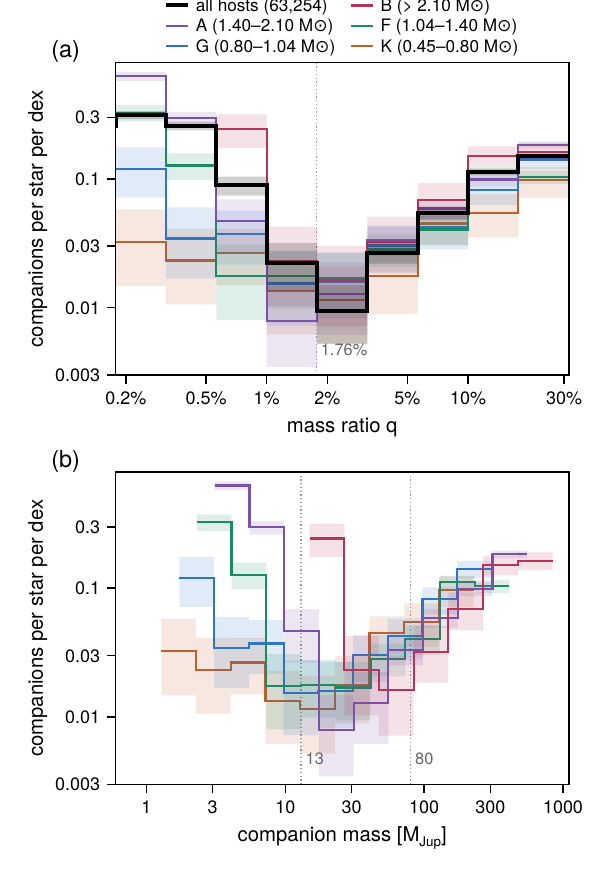}
\caption{Companion rates versus mass ratio at 3--94\,yr in five host-mass ranges. The lines and shaded bands are posterior medians and 16--84th percentiles. The desert minimum sits at $q \approx 2\%$ in every range, and the stellar side (right) is similar across host masses. The solid black line in panel (a) is the full sample of 63,254 stars, masked as in Figure~\ref{fig:funnel}. Panel (b) shows the same profiles against companion mass (mean host mass per range); dotted lines mark 13 and 80\,\mjup{}.}
\label{fig:desert}
\end{figure}

\section{Results and Discussion}\label{sec:results}

Figure \ref{fig:funnel} shows the posterior median over the full sample of 63,254 stars on the expected number of companions $/$ star $/$ dex of period $/$ dex of mass ratio (that is, $\;\bar N\,\pi_\ell/(\Delta\log P\,\Delta\log q)$, where $\bar N=\sum_n n\,\psi_n\;$; bin edges are listed in \S\ref{sec:data}). We mask cells where fewer than 300 stars are sensitive (95th-percentile posterior $q$ below the cell), where the prior dominates.
Our rates should be read as expected number of companions of that $(\log{P}, \log{q})$ around stars, rather than the fraction of systems that are multiple.

We find two detectable populations of companions: an upper band of companions with mass ratios $q=1 - 0.08$ (stellar around most stars; BDs around lower mass stars) with periods ranging from a few days to $10^5\,\rm{yr}$. Close stellar companions are detected with an occurrence rate that drops off sharply near $\sim3$ days. The number of near unity mass-ratio stellar companions drops beyond $\sim$540 years due to the sample selecting against Hipparcos non standard solution types. The rates and distribution vs. period are comparable to those reported by \citet{moe_distefano_2017} (see Appendix \ref{ap:known}).
Below a few percent mass-ratio, we find a second population of low mass-ratio companions, with the best sensitivity in our 3--94\,yr bins. The rate of low mass-ratio planets is significantly higher near 10\,yr  than inwards of 0.5\,yr---consistent with previous findings on the rarity of hot-Jupiter and super-Jovian companions \citep{wright_2012,cumming_2008}.

\paragraph{The desert extends to $\sim100$ yr periods around all stars} In between these two evidently separate populations of companions at 3--94\,yr lies a mass-ratio desert, with a significant deficit in occurrence rate compared to the two bounding populations of companions. In the full-sample fit, the occurrence at the floor of the desert ($q = 1.8$--3.2\%) is lower by a factor of $33^{+26}_{-12}$ than at the planet-side peak and by $19^{+15}_{-7}$ than at the stellar-side peak.
It should be noted that the prior we adopted on the companion distribution is towards globally flat occurrence, away from the presence of features like the mass-ratio desert (Appendix \ref{ap:priors}).
Outwards of $\sim$100\,yr, we cannot determine if the desert persists due to fundamental degeneracy between companion mass and separation in the Hipparcos-Gaia acceleration regime.
Targeted imaging follow-up of Hipparcos–Gaia accelerators \citep[e.g.][]{currie_2023} combined with joint astrometry and imaging demographics would likely be able to split this population (see Appendix \ref{ap:gpies} for a direct comparison with GPIES).

\paragraph{The desert is a feature in mass-ratio across host mass}\label{sec:mass}

We compared the occurrence rate of companions with orbital periods between 3 and 94\,yr for hosts across 0.16–8.1 \msun{} (99\% lying between 0.54–5.1 \msun{}).
In Figure~\ref{fig:desertmap}, we plot the 3--94\,yr columns of posteriors fit to six host mass ranges side by side, in units of companion mass versus host-mass.

To determine how the minimum moves with host mass, we find the minimum of the desert in each host-mass range of Figures \ref{fig:desertmap} and \ref{fig:desert} by fitting a quadratic in mass-ratio, and then fit a simple constant ratio to these mass-ratio minima. We find that the minimum is at $q=1.76^{+0.26}_{-0.19}\%$ ($\chi^2 = 1.6$ for 4 degrees of freedom) across the sensitivity-selected K--B stars (mean host masses 0.69--2.54\,\msun{}). A single fit to all 63,254 hosts places the minimum at $q=2.31^{+0.13}_{-0.39}\%$, consistent with this line at $1.2\sigma$ (and with the rates at the floor of the desert at $0.8\sigma$). We exclude stars below 0.45\,\msun{} from this fit due to the small sample size (119 stars only), but do see evidence (Figure \ref{fig:desertmap}) of the trend continuing. This would be consistent with the findings of \citet{zhang_2025_desert} which found a break in the planet population near $q\approx0.02$ around low-mass stars. The scaling of the desert with mass ratio is unambiguous in these results as companions above the mass-ratio desert on lower mass stars essentially overlap in mass with companions below the mass-ratio desert on the highest mass stars in our sample (Figure \ref{fig:desert}(b)).

\paragraph{Opposite metallicity trends either side of the desert}\label{sec:metallicity}

\begin{figure}[t]
\centering
\includegraphics[width=\columnwidth]{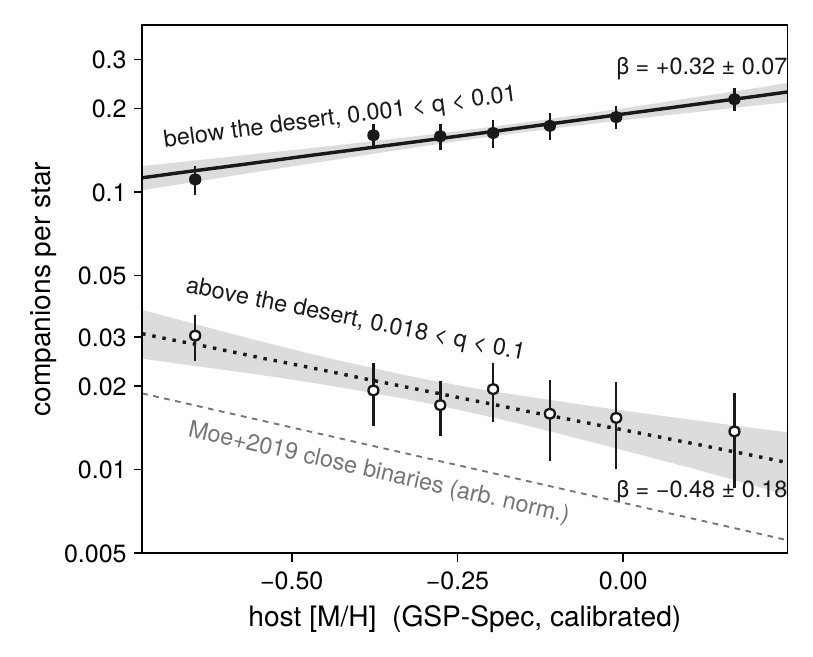}
\caption{Occurrence rate versus metallicity for companions above and below the 2\% mass-ratio desert for FGK hosts at 3--94\,yr (18,672 FGK hosts). While low mass-ratio companions occur more often with increasing metallicity, companions above the mass-ratio desert occur less often. This supports the picture that the desert lies between two different formation channels, each scaling with host-mass. The dashed line shows the slope of the close ($a < 10$\,au, $q > 0.1$) binary fraction from \citet{moe_2019}, with arbitrary normalization.}
\label{fig:feh}
\end{figure}

We compared the occurrence rate of companions below and above the 2\% mass-ratio valley as a function of metallicity (Figure \ref{fig:feh}). We fit the joint population model to hosts in seven [M/H] bins with equal numbers of stars. Because the per-system posteriors are diffuse, we adopt a two-step Bayesian update for this fit. Each bin's fit uses as its prior the posterior mean of a flat-prior fit to all other stars (leave-one-out, so no data is counted twice), which borrows the shape in $\log P$ and $\log q$ while assuming no trend with \feh{}.

We then fit a power law, occurrence $\propto 10^{\beta\feh}$, to these seven posteriors. Around FGK stars, companions between 0.1\% and 1\% mass-ratio, below the desert, have a positive exponent with $\beta=+0.32 \pm 0.07$. A flat trend is rejected at $4.4\sigma$.
By contrast, companions above the desert between 1.8\% and 10\% mass-ratio have a negative exponent $\beta=-0.48 \pm 0.18$. A flat trend above the desert is disfavored by $2.6\sigma$.
Using the California Legacy Survey, \citet{giacalone_2025} independently find that host metallicity changes at a companion mass of $27^{+12}_{-8}\,\mjup$ for companions at 1--50\,au, which corresponds to $q\approx2.6\%$ for a solar-mass host, close to the desert we find.

The positive exponent for low mass-ratio companions is expected for formation through core accretion, but it is shallower than the planet--metallicity relation found by RV surveys for giant planets on $\lesssim4$\,yr orbits, where occurrence scales as $10^{2.0\,\feh}$ \citep{fischer_valenti_2005} or $10^{1.2\,\feh}$ \citep{johnson_2010}. Part of this difference is likely astrophysical: our window extends to $\sim20$\,au and up to $q=1\%$ ($\sim10\,\mjup$ for a Sun-like host), and both wide-orbit imaged planets and giant planets above $\sim4\,\mjup$ show a weaker preference for metal-rich hosts \citep{santos_2017,swastik_2021}.  Gaia DR4's improved period constraints may help separate these effects. Stellar companions are known to have occurrence that decreases somewhat with metallicity \citep{moe_2019}, just as we see in the above-desert companion rate. Previous work has shown that hosts of companions in the brown-dwarf desert likewise show weak or no metallicity enhancement \citep{ma_ge_2014,maldonado_villaver_2017,schlaufman_2018}, in line with our negative above-desert slope.

\paragraph{Agreement with other samples and studies} A unique property of this study is that we consider both stellar and planetary companions in a single framework. We test that fits to stars selected to have known stellar companions (1,579), brown dwarfs (53) and Jovian planets (201) recover them at reasonable periods and mass ratios (Appendix~\ref{ap:surveys}).

In Appendix~\ref{ap:stellar}, we examine the stellar companion occurrence rates as a function of $\log{P}$, and find qualitatively good agreement and rates mostly between  \citet{moe_distefano_2017} and \citet{raghavan_2010}. The exception is an elevated stellar companion rate in the longest period bin, approaching $10^5$ yr, which appears to be dominated by chance alignments and does not impact the values we quote at 3--94\,yr.

Restricted to the stars that RV surveys targeted, we find the desert minimum at a similar companion mass to \citet{cui_2026} on a similar sample (about 0.2\,dex lower), though at lower overall occurrence (Appendix~\ref{ap:cui}). For works that quoted a specific target list, an even more direct test is possible by fitting our model to their targets. Our planet rates agree with the California Legacy Survey \citep{Rosenthal2021} on the 622 of its stars in our sample where our sensitivities overlap (Appendix~\ref{ap:cls}), and with GPIES \citep{nielsen_2019} on the 238 of its stars in our sample, overall and versus host mass (Appendix~\ref{ap:gpies}).

\paragraph{Robustness} In Appendix \ref{ap:robust}, we consider if our results show sensitivity to systematics, particularly miscalibrated astrometric uncertainties as functions of color (main sequence vs.\ giant branch) and scan law (sky position bins). We see no evidence of color or position dependent changes to the low mass-ratio companion occurrence rate.

\paragraph{The desert with orbital period and eccentricity}

Pre-DR4 data do not constrain the period of individual planets well, so Figure~\ref{fig:funnel} does not confirm that planets extend to 100\,yr with flat occurrence. On the other hand, companions above the desert have better orbital period constraints and hint that the upper limits of the desert may move upwards in mass-ratio with orbital period (Figure \ref{fig:funnel}, across all host masses). 

Previous studies have shown distinct eccentricity distributions above and below the BD desert among certain host mass ranges \citep{bowler_blunt_2020_ecc,DoO_2023_ecc,Gilbert_2026}. The data used here pre-DR4 are insufficiently constraining to measure this difference.

\section{Summary and Conclusion}\label{sec:conc}

This work has presented the first population-level planet demographics with astrometry as the primary constraint. Using Gaia DR2, DR3, and Hipparcos data for 63,254 stars, we find that the brown dwarf desert is a feature in mass ratio, with a minimum at  $q=1.76^{+0.26}_{-0.19}\%$ for hosts of 0.54--5.1\,\msun{} (99\% of the sample; Figures \ref{fig:funnel}--\ref{fig:desert}). The companion occurrence rate scales positively with metallicity below 1\% mass-ratio and negatively between 1.8\% and 10\% (Figure \ref{fig:feh}). This is consistent with the desert being the gap between two formation channels that both scale with host mass and scale oppositely with metallicity. Companions are detected below the mass-ratio desert with rising occurrence vs.\ host mass, showing that wide outer companions continue to be formed around more massive stars than previous surveys have accessed. At their most extreme, these low mass-ratio companions exceed the $\sim12.7\,\mjup$ deuterium burning limit yet appear to be a smooth continuation of the planet population from Sun-like stars.

\paragraph{Outlook} With the approaching release of Gaia DR4 epoch astrometry, we expect this mass-ratio desert to sharpen considerably. DR4 will constrain below-desert companions for many more stars, especially lower-mass hosts, and measure periods and eccentricities that may reveal trends in the desert with separation or eccentricity \citep[e.g.][]{DoO_2023_ecc}. That said, below-desert companions with orbital periods outwards of $\sim5$ yr will not be well characterized by DR4 alone, and would still benefit from including data from Hipparcos, where available. The modeling approach developed here and in \paperII{} will be applicable to DR4 epoch astrometry.

\begin{acknowledgments}

This work has made use of data from the European Space Agency (ESA) mission Gaia (\href{https://www.cosmos.esa.int/gaia}{https://www.cosmos.esa.int/gaia}), processed by the Gaia Data Processing and Analysis Consortium (DPAC),\\\href{https://www.cosmos.esa.int/web/gaia/dpac/consortium}{https://www.cosmos.esa.int/web/gaia/dpac/consortium.} This work made use of the Gaia Observation Forecast Tool (GOST). This research used the facilities of the Canadian Astronomy Data Centre operated by the National Research Council of Canada with the support of the Canadian Space Agency.
This research was enabled in part by support provided by the Digital Research Alliance of Canada (alliancecan.ca).
The Pigeons package's development and collaboration are supported by a Canadian Statistical Sciences Institute Collaborative Research Teams Grant.

J.S. acknowledges financial support from the Heising–Simons Foundation through the 51 Pegasi b Fellowship (grant \#2025-5885).
J.W.X is thankful for support from the Heising-Simons Foundation 51 Pegasi b Fellowship (grant \#2025-5887).
D.J.\ is supported by NRC Canada and by an NSERC Discovery Grant.

This work benefited from the 2025 and 2026 Exoplanet Summer Program in the Other Worlds Laboratory (OWL) at the University of California, Santa Cruz, a program funded by the Heising-Simons Foundation and NASA.
\end{acknowledgments}

\section*{Data availability}
The per-star posteriors underlying this work are released with \paperII{} and will be made publicly available through the Canadian Astronomy Data Centre upon acceptance; they can already be browsed at \url{https://exopo.st/}. The analysis code and the population posteriors will likewise be made publicly available upon acceptance.

\software{Octopodes.jl, Octofitter \citep{octofitter}, Pigeons.jl \citep{pigeons}, Julia \citep{julia_bezanson}}

\bibliography{ms}{}
\bibliographystyle{aasjournalv7}

\appendix
\input{app-3}

\input{app-1}
\input{app-2}

\end{document}

%% file: app-3.tex
\section{Recovery of known populations and surveys}\label{ap:surveys}

The methods used in this Letter differ from the majority of demographics studies. The common approach is to start from a vetted sample and a list of detections, each with a well-measured period and mass (or $m\sin i$, or radius), and then correct for completeness with injection--recovery tests.
The per-star posteriors of \paperII{} are far more ambiguous than RV or transit detections: most are at low signal-to-noise, many span several decades in period, and for many stars the data alone cannot tell a stellar companion from a planet. We instead keep every star, with no binary vetting, and sensitivity enters through each star's likelihood rather than through a completeness map.
As such, the results we present rely more heavily on the joint population model and large sample size.
The population model separates sub-populations of companions by pooling information across stars, and the result differs markedly from a simple stack of the per-star posteriors. This has four consequences that may be unfamiliar.
\begin{itemize}
\item \textbf{Weakly constrained companions can still produce a sharp population.} Each star's posterior might, e.g., spread along a ridge of roughly constant $m/a^2$ or constant RV semi-amplitude $K$, but these ridges from different stars cross. A population concentrated where they overlap explains the observations of all stars at once, whereas a population spread along every ridge is a more complex solution. By integrating over the population posterior, the concentrated solution is favored. This allows a compact peak of giant planets to emerge from posteriors that individually span decades in period and mass.
\item \textbf{Nearby stars inform the population.} Around nearby stars, stellar companions are reliably detected even with near-equal flux (see \paperII{}, \S3.3.5). This is because any low-contrast stellar companion can be detected either through excess Gaia RV variance or resolved as a separate source. Because the population model assumes exchangeability---that nearby stars host the same distributions of companions as distant stars---the population model learns about these companions even around distant stars whose data alone cannot distinguish them from planets, and vice versa.
\item \textbf{The rate in a region is not a count of detected companions.} Through the shared population, every star affects every region to some degree. Rates in separate regions of the $(\log P, \log q)$ parameter space are properties of the population fit and cannot be attributed to a particular companion around a particular star.
\item \textbf{Fits to smaller samples become uncertain in where a population lies, not just in the companion rates.} With a few hundred stars there may be too little information to split degenerate populations the way the full population fit does.
Adding more stars to the sample increases the concentration of the population posterior upwards and downwards away from the desert.
\end{itemize}

This approach is very powerful in its ability to pool information from stars without strong, well-constrained detections. Because it departs from standard practice, we check it here against cases with known answers.
First, we refit the population on hosts with cataloged companions: visual (ORB6) and spectroscopic (SB9) binaries, single stars with known cold Jupiters, hosts of known brown dwarfs, and, as a null control, survey targets with no detected companion (Appendix~\ref{ap:known}). As we will show, the population posteriors recover $(\log P, \log q)$ distributions consistent with known objects from external catalogs.
The samples used in the main Letter, by virtue of their larger sample sizes, can be expected to concentrate even more strongly towards distinct sub-populations.
Second, we compare our rates with independent demographics results: stellar multiplicity from \citet{raghavan_2010} and \citet{moe_distefano_2017} (Appendix~\ref{ap:stellar}), RV plus astrometry from \citet{cui_2026} (Appendix~\ref{ap:cui}), RV from the California Legacy Survey (Appendix~\ref{ap:cls}), and imaging from the Gemini Planet Imager Exoplanet Survey (Appendix~\ref{ap:gpies}). For CLS and GPIES, we fit our population model to the \emph{same} stars used in those surveys for a like-to-like comparison. We find rates and distributions that are consistent with known binary rates and period distributions, and companion rates consistent with CLS and GPIES on those surveys' own stars. For \citet{cui_2026}, we recover a consistent location of the brown dwarf desert in companion mass around their target mix, but with lower overall occurrence than their reported completeness-corrected results.

\begin{figure*}[p]
\centering
\includegraphics[width=\textwidth]{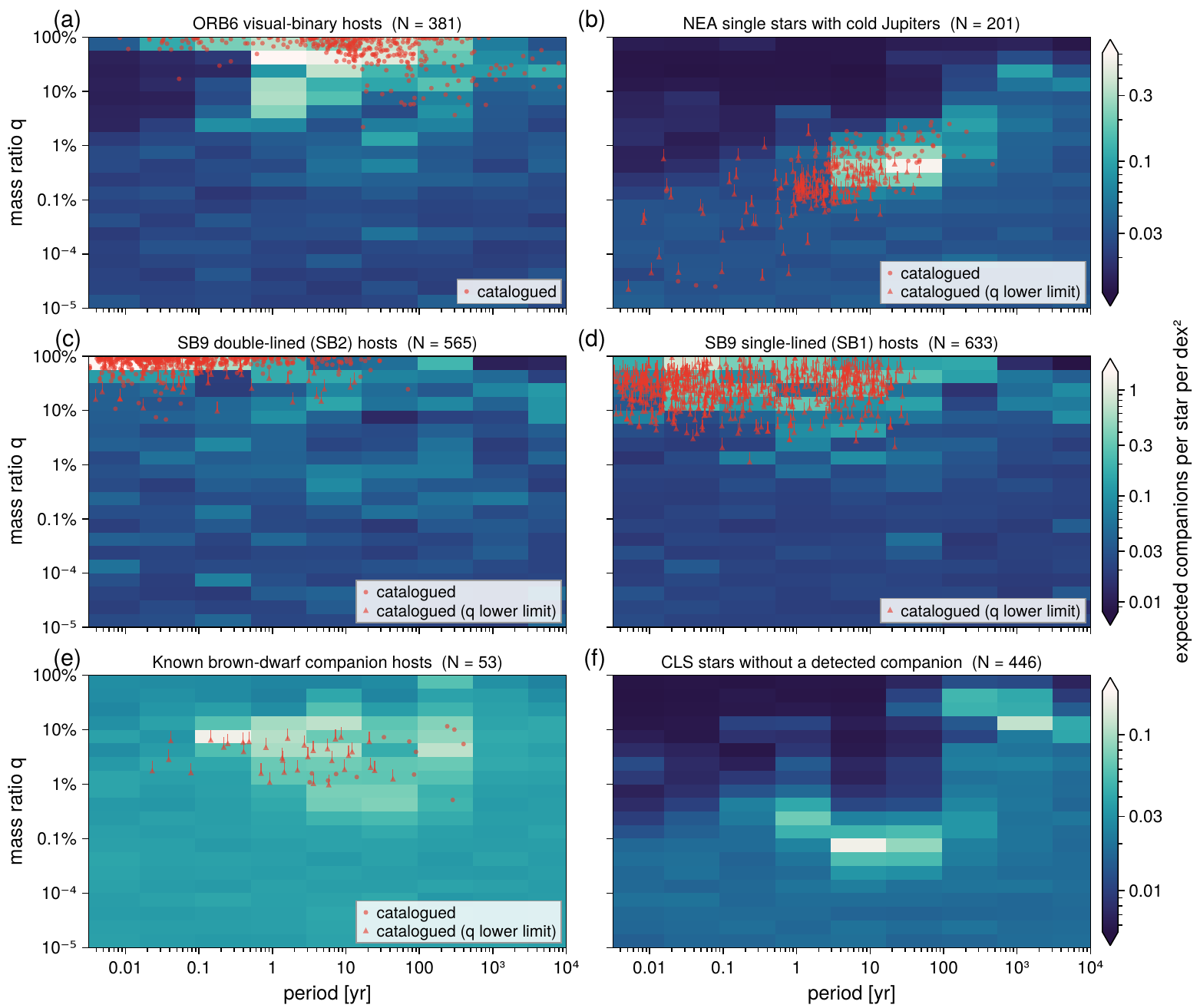}
\caption{Population fits to sets of hosts with known companions. Each row has its own color scale. (a) hosts of visual binaries with a graded (1--5) orbit in the ORB6 catalog \citep{orb6}. (b) single stars (one star in the system) in the NASA Exoplanet Archive \citep{Christiansen2025} hosting a planet of more than 1\,\mjup{} (true mass or $m\sin i$) on a period longer than one year. The binaries populate the stellar band at $q > 0.1$, with a tail to $q \approx 0.01$; the cold-Jupiter hosts produce a compact peak at $q \approx 3\times10^{-3}$--$10^{-2}$, 3--100\,yr, and no stellar companions inside $\sim$100\,yr. (c) hosts of double-lined (SB2) spectroscopic binaries in SB9 \citep{Pourbaix2004SB9}, with $q = K_1/K_2$. (d) single-lined (SB1) systems in SB9, with the minimum $q$ from the mass function (lower limits). (e) hosts of known brown-dwarf companions (13--80\,\mjup{}): RV companions from \citet{ma_ge_2014}, \citet{kiefer_2019} and the NASA Exoplanet Archive ($m\sin i$; lower limits), and directly imaged brown dwarfs with dynamical masses (circles). (f) a null control on the BD regime: the targets of the California Legacy Survey (CLS) with no detected companion. A very low occurrence of mid mass-ratio companions is recovered as desired. An additional population of low mass-ratio companions is recovered near the detection limits of the CLS, expected since the CLS completeness drops to $\approx1/3$ near $q=10^{-3}$ at these periods, so that $\approx2/3$ of such companions are missed. Red markers are the cataloged companions, with $q$ computed from our host mass: for ORB6 from the magnitude difference, $q = 10^{-\Delta m/8.75}$; where the mass is $m\sin i$ (NEA, SB1, RV brown dwarfs) the companion is drawn as an upward triangle. The uniform floor is the flat Dirichlet prior on occurrence in absence of data. Populations consisting mainly of stellar companions ((a), (c), (d)) do not have sensitivity to lower mass companions.}
\label{fig:known}
\end{figure*}

\subsection{Known binaries and known cold Jupiters}\label{ap:known}

As a check that stellar binaries and giant planets are separated and that binaries do not leak into the planet regime of Figure~\ref{fig:funnel}, we refit the population model on two sets of hosts with known companions (Figure~\ref{fig:known}(a)--(b)). For 381 ORB6 visual-binary hosts in our sample, the model returns 1.16 companions per star at $0.1 < q < 1$, mostly concentrating in an upper band in mass-ratio, as expected.
For 201 single stars with known cold Jupiters our model returns 0.39 per star in the Jovian window. The recovered peak lies on the cataloged planets at 3--100\,yr. The cataloged planets at $\sim$1\,yr and $q \lesssim 10^{-3}$ (mostly $m\sin i$ from RV) fall below our sensitivity, and the population stays close to the occurrence rate prior there.

The spectroscopic binaries test the short-period corner of Figure~\ref{fig:funnel}. For 565 SB2 hosts the model concentrates strongly in the top-left corner of Figure~\ref{fig:known}(c). For the 633 SB1 hosts, the posterior is spread over $q \approx 0.1$--1 above the cataloged minimum mass ratios from the catalog. Neither set leaks strongly into the planet regime.

Known brown dwarfs in Figure~\ref{fig:known}(e) test the model's ability to detect companions in the mass-ratio desert itself. The sample size is small (just 53 hosts) so the population model is limited in its ability to concentrate in ($\log{P}, \log{q})$. Still, the BDs are recovered as a distribution of companions at intermediate mass-ratios and periods.
As a null test, the 446 California Legacy Survey targets in our sample without any detected RV-companion recover no such occurrence of BDs. A notable occurrence bump at $q \approx 6\times10^{-4}$--$10^{-3}$ and 3--30\,yr is seen in our results, near the survey's detection limits. These are the planets the CLS missed due to its $\approx1/3$ completeness in that part of the parameter space. A quantitative comparison with CLS occurrences is in Appendix~\ref{ap:cls}.

\subsection{Stellar companions}\label{ap:stellar}

\begin{figure*}
\centering
\includegraphics[width=0.6\textwidth]{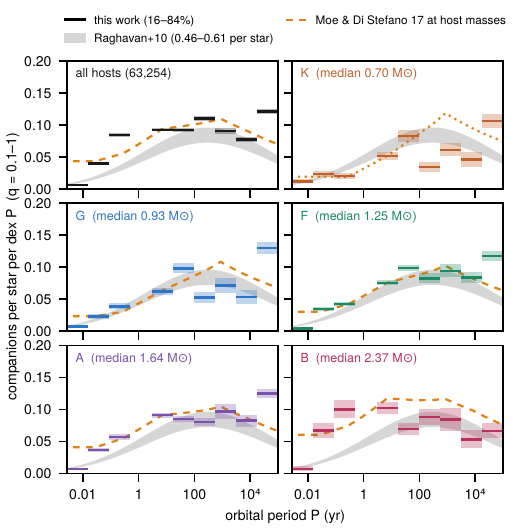}
\caption{Comparison of stellar ($0.1 < q < 1$) occurrence rates vs. orbital period compared against \citet{raghavan_2010} and \citet{moe_distefano_2017}. The 0.5--2.9\,yr column is omitted (one-year alias). Gray band: the log-normal period distribution of \citet{raghavan_2010} ($\mu = 5.03$, $\sigma = 2.28$ in $\log P$\,[d]), scaled to 0.46--0.61 companions per star (their multiplicity fraction and companion frequency) and by 0.9 for $q > 0.1$. Orange: $f_{\log P;\,q>0.1}$ of \citet{moe_distefano_2017} averaged over each panel's host masses (dotted for K, below their 0.8\,\msun{} calibration).}
\label{fig:stellarcomp}
\end{figure*}

We now check that our inferred population of \emph{stellar} companions tracks the known rate and distribution of binary stars versus period. In Figure \ref{fig:stellarcomp}, we show the companion rate we infer from Gaia and Hipparcos from 1 d to $10^5\,$yr (the prior bounds of \paperII{}) for the full population, and separated for K, G, F, A, and B stars. The 1 yr column is skipped. We find a distribution of companions that qualitatively matches \citet{moe_distefano_2017} well, and is similar to the log normal distribution presented in \citet{raghavan_2010}. The results of \citet{moe_distefano_2017} required synthesizing occurrences from  a compilation of surveys made with at least five techniques (spectroscopy, eclipses, long-baseline interferometry, adaptive optics and other high-resolution imaging, and common proper motion), each sensitive to only a narrow range of P and q, and correcting each for its own selection effects. Remarkably, our model recovers comparable distributions from a single, homogeneous all-sky dataset.

The major source of disagreement is in the widest period bin, which is informed almost entirely in our models by common proper motion companions, and which appears at consistently higher rates than other studies. These are likely dominated by chance alignments at periods where astrometry has no constraining power to test consistency of the candidate companions' kinematics with observed reflex motion. The addition of differential Gaia RVs between the host and common proper motion candidates may help to resolve this in future (discussed briefly in \paperII{}). For the purposes of this Letter, an elevated rate of very wide companions does not impact the presented 3--94\,yr occurrence rates.

\subsection{RV + astrometry: \citet{cui_2026}}\label{ap:cui}

\begin{figure*}
\centering
\includegraphics[width=\textwidth]{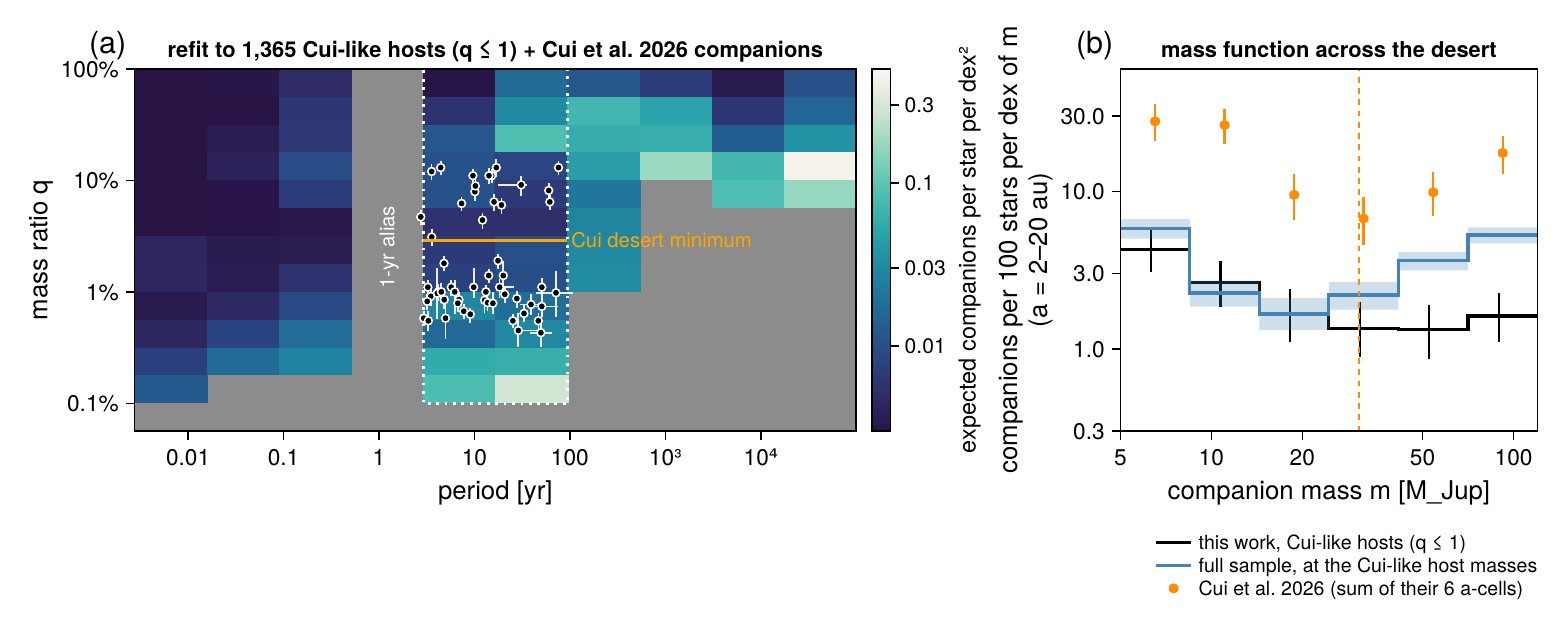}
\caption{Comparison with \citet{cui_2026} (RV and astrometry of RV-detected hosts). (a) our inferred companion rates on an approximation of the target list of \citet{cui_2026}, with their detections overplotted (their Supplementary table; median and 68\% interval in $P$ and $q$). Their 790-star target list is unpublished, so this sample is an approximate reconstruction: the CLS and HARPS GTO targets with their cuts (0.6--1.4\,\msun{}, $\log g > 3.9$). The orange line is the desert minimum they report, $30.9\,\mjup{}$, at the median host mass implied by their table (1.02\,\msun{}).  (b) slice through companion mass with both results compared. The black series traces the fit of the left panel, while the blue series is from our full sample (Figure \ref{fig:desert}), evaluated at Cui-like host masses (blue, 68\% band). The desert falls at a similar companion mass, though the normalization is significantly different, with \citet{cui_2026} reporting 3--11$\times$ more companions in every mass bin. This difference might be attributable to the completeness correction made in that work, as the overall rates we find are in good agreement with other surveys (Appendices~\ref{ap:cls} and \ref{ap:gpies}). }
\label{fig:cui}
\end{figure*}

\citet{cui_2026} combined radial velocities with Hipparcos--Gaia astrometry for 790 FGK stars (0.6--1.4\,\msun{}) from four underlying RV surveys and found 55 companions of 5--120\,\mjup{} at 2--20\,au. This work is perhaps the most directly relevant point of comparison with this Letter. They place the desert minimum at $30.9^{+5.5}_{-4.8}$\,\mjup{}, i.e.\ $q_0 = 0.029$ at their median host mass, and measure completeness-corrected rates of $12.6^{+2.3}_{-1.9}\%$ for 5--14\,\mjup{}, $1.6^{+0.7}_{-0.5}\%$ for 24--42\,\mjup{} and $4.2^{+1.2}_{-1.0}\%$ for 71--120\,\mjup{}. Their 2--20\,au range at 1\,\msun{} is essentially our 3--94\,yr window (Figure~\ref{fig:cui}(a)), and our desert minimum, $q_0 = 0.018^{+0.003}_{-0.002}$ (Section~\ref{sec:results}), lies $0.21$\,dex (about $2\sigma$) lower; in our G stratum alone the minimum sits at $17.1^{+6.2}_{-3.8}$\,\mjup{}.

The location of the desert broadly agrees, but the normalization does not (Figure~\ref{fig:cui}(b)). Our full-population profile at Cui-like host masses reaches its minimum in the 14--24\,\mjup{} bin, one bin below theirs at 24--42\,\mjup{} and consistent with the 0.2\,dex offset above, and rises again towards 71--120\,\mjup{} as theirs does; the planet-to-desert contrast in their bins is 3.7 for ours and 7.9 for theirs. Our rates are lower in every bin: the refit to the Cui-like stars gives $1.6^{+0.5}_{-0.4}\%$ for 5--14\,\mjup{}, $0.31^{+0.14}_{-0.11}\%$ for 24--42\,\mjup{} and $0.38^{+0.15}_{-0.12}\%$ for 71--120\,\mjup{}, factors of 8, 5 and 11 below \citet{cui_2026}, and $3.0 \pm 0.6\%$ in total for 5--120\,\mjup{} (4.8\% from our full population at the same host masses), against their $23.1\%$. On our approximation of their target list, which contains 38 of their companion hosts among 1,365 stars (2.8\%), the refit total matches the raw detection fraction. Their own raw detection fraction is 55/790 = 7\%, so the difference may lie in their factor of $\sim$3.3 completeness correction.

Their presented rate of low mass companions is also high compared with RV alone. In the CLS non-parametric posterior \citep{fulton_2021}, all companions of 1.3--19\,\mjup{} ($M \sin i$) at 2.4--16\,au together amount to $4.2^{+1.0}_{-0.8}$ per 100 stars, three times below the $12.6\%$ that \citet{cui_2026} find for the narrower 5--14\,\mjup{} range at 2--20\,au ($9.8\%$ from their cells inside 2--13.6\,au). Scaling by the share of 5--14\,\mjup{} in the CLS mass function at 1--5\,au gives $\sim$$1.2\%$ for that mass range. On the CLS stars our population gives $1.6^{+0.1}_{-0.2}\%$ for 5--14\,\mjup{} at 2--20\,au in $M \sin i$, close to the CLS value, so the tension is between \citet{cui_2026} and both other measurements. The boxes do not match exactly (true mass against $M \sin i$, 2--20 against 2.4--16\,au, and different host samples), but none of these differences is sufficient to explain a factor of three, and their completeness correction is a plausible source of the disagreement.

\subsection{RV: the California Legacy Survey}\label{ap:cls}

\begin{figure*}
\centering
\includegraphics[width=\textwidth]{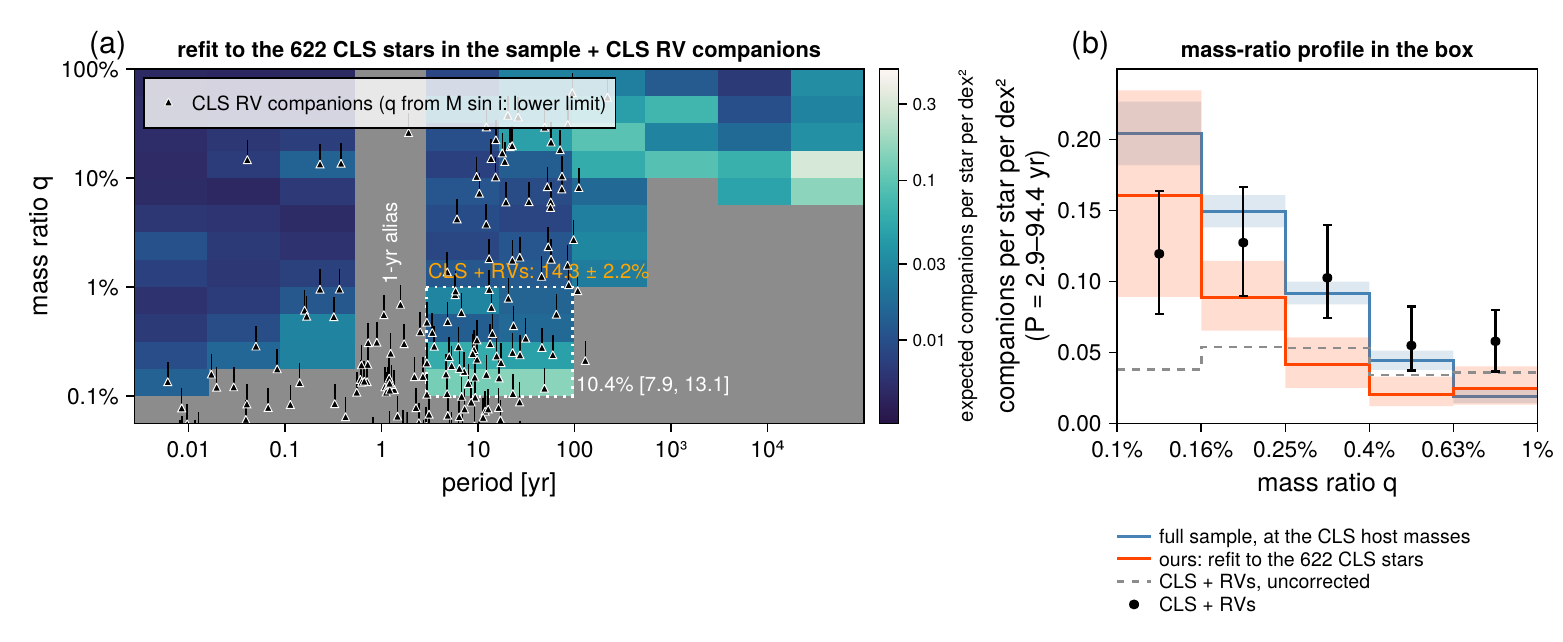}
\caption{Comparison with \citet{Rosenthal2021} (The California Legacy Survey, or CLS). (a) A fit to the 622 (of 719) CLS targets in our sample, with the box $10^{-3} < q < 10^{-2}$, $3 < P < 94$\,yr and our rate inside it, and the CLS RV companions on those stars ($q$ from $M\sin i$ and our host masses; lower limits drawn as upward triangles). Orange: the completeness-corrected CLS RV occurrence rate in the box. Gray cells are masked as in Figure~\ref{fig:funnel}. (b) Companions per star per dex$^2$ (of $\log q$ and $\log P$) in that box. Points: CLS + RVs, completeness corrected; dashed gray: the same without the completeness correction. Orange: the occurrence from the left panel. Bands are 68\%. Blue: our full population fits weighted to the host mass distribution of this sample.}
\label{fig:cls}
\end{figure*}

In Figure \ref{fig:cls}, we compare the occurrence rate as a function of mass-ratio we infer from a set of 622 specific targets shared by both our sample and the CLS. By using the exact same targets, we can be confident that no selection effects impact the comparison.

In the box $10^{-3} < q < 10^{-2}$ and $3 < P < 94$\,yr, our inferred rate is $10.4^{+2.7}_{-2.5}$ companions per 100 stars. We also consider our full-population astrometry results (Figure \ref{fig:desert}), weighted by the host masses of the CLS. This gives $15.4^{+0.8}_{-0.9}$.

We then performed an independent analysis of the same 622 stars using the sample and occurrence methodology of \citet{VanZandt_2026}. They combined the CLS RVs with Hipparcos-Gaia astrometry, and applied an isotropic $\sin i$ distribution where no astrometric solution existed. Their hierarchical Poisson occurrence model incorporates companion posteriors and injection/recovery maps to correct for missed planets. Using this approach, we found $14.3 \pm 2.2$ companions per 100 stars in this box. Both our per-bin and summed rates using these two methods agree within their respective errors (Figure~\ref{fig:cls}(b)),
but the shapes differ slightly, with our rates showing a steeper drop off in occurrence with mass-ratio. The overall rates remain compatible.

Separately, using the full 719-star CLS sample with Hipparcos--Gaia true masses, \citet{VanZandt_Bowler_2026} find that the brown dwarf desert begins at $q=1.3$--$2.0\%$, compatible with the desert minimum we find, and that the desert is more sharply defined in mass ratio than in companion mass. They also find that the mass-ratio functions of hosts above and below $1\,\msun{}$ agree within $1.6\sigma$, while their true-mass functions differ at about $2.4\sigma$, although over a narrow range in host mass.

Figure~\ref{fig:known}(f) is also of interest: this is our fit to only the CLS targets without confirmed detections. A population of planetary companions is seen near $q=10^{-3}$. It is in fact expected and consistent with the CLS demographics results that we should find additional companions around these stars, as their $\approx1/3$ completeness near $q=10^{-3}$ implies that $\approx2/3$ of such companions go undetected. These non-detections would be correlated with those of other RV surveys (searches would fail for correlated reasons e.g. according to the inclination of the companion's orbit), but not with an independent detection technique like astrometry. As such, an independent method applied to targets selected on RV non-detections is expected to still detect companions.

\subsection{Imaging: GPIES}\label{ap:gpies}

\begin{figure*}
\centering
\includegraphics[width=\textwidth]{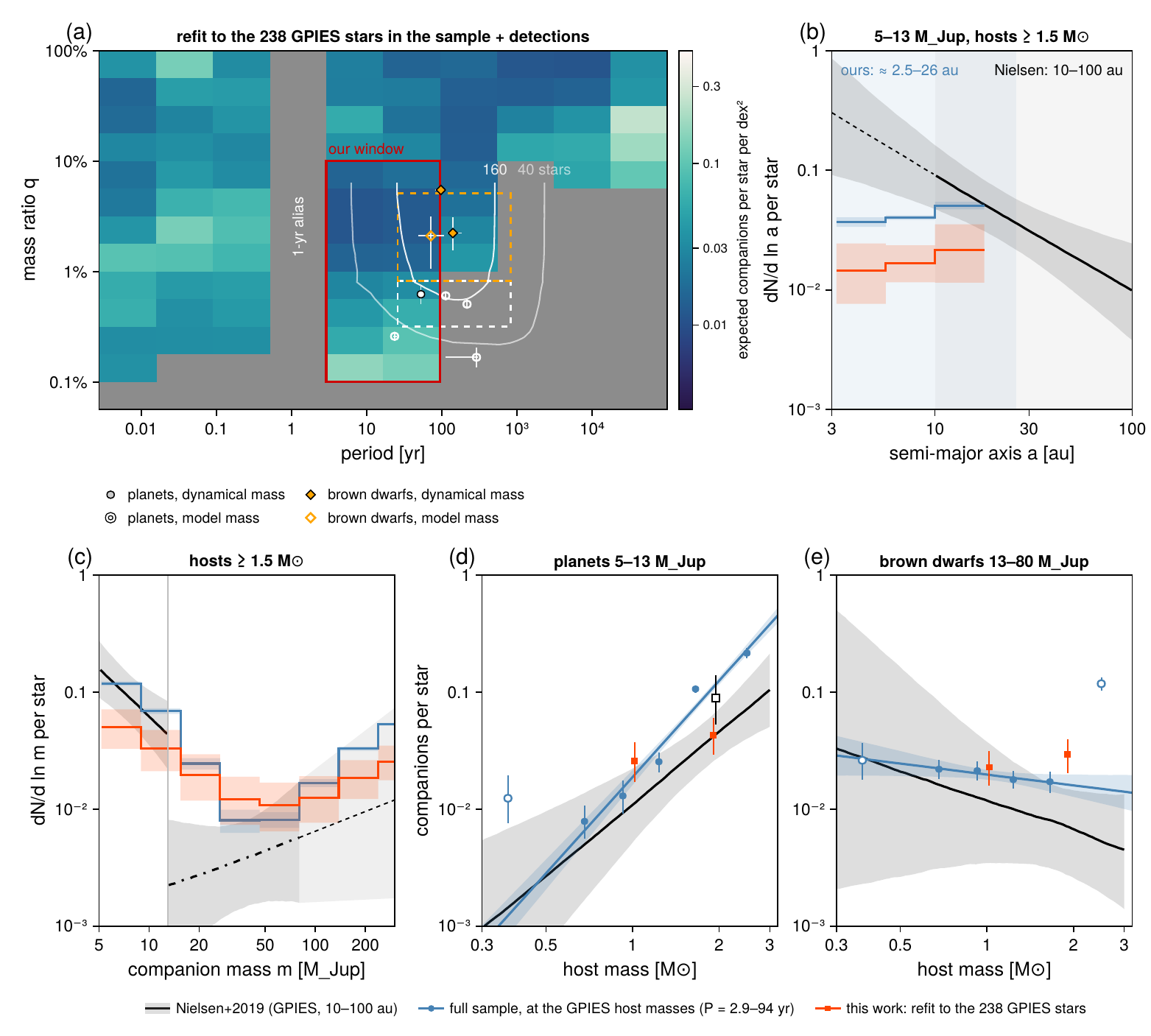}
\caption{Comparison with GPIES \citep{nielsen_2019}. (a) A population fit to the 238 GPIES targets in our sample, with the GPIES detections around them at their published masses and orbital periods (filled: dynamical masses; open: luminosity-based model masses; sources in the text), the boxes of \citet{nielsen_2019} (5--13\,\mjup{}, white; 13--80\,\mjup{}, orange; both 10--100\,au) and the GPIES depth-of-search contours (40 and 160 of 300 stars complete), mapped to $(P, q)$ at 1.5\,\msun{}. Red: the period range of our rates in (b)--(e), $P = 3$--94\,yr. (b) $dN/d\ln a$ of 5--13\,\mjup{} companions for the 97 targets above 1.5\,\msun{}; shading marks the two separation ranges. (c) $dN/d\ln m$ for the same hosts, with the GPIES brown-dwarf law (dash-dotted) extrapolated beyond 80\,\mjup{} (thin dashed). (d, e) Companions per star against host mass for 5--13 and 13--80\,\mjup{}; open square: the GPIES $> 1.5$\,\msun{} planet rate. Throughout, black and gray are \citet{nielsen_2019} (10--100\,au); blue is our full-sample population weighted to these hosts' masses (points: host-mass strata; line: power-law fit to the filled points; open circles are not fit); orange is our refit to the GPIES stars. Bands are 68\%.}
\label{fig:gpies}
\end{figure*}

We finally compare the occurrence rates we infer against those from the Gemini Planet Imager Exoplanet Survey \citep[GPIES;][]{nielsen_2019} in Figure \ref{fig:gpies}. We compare the ``tongue plot'' of \citet{nielsen_2019} with our population model fit to 238 of the 300 specific targets in the GPIES.
As with the CLS comparison in the preceding section, by using (nearly) the same specific targets, we can be confident that no selection effects between the two impact the comparison. Looking at Figure \ref{fig:gpies}(a), planetary companions are observed below $q=0.01$, and a dearth of moderate-separation stellar companions is found compared to our overall population fits, owing to the campaign's binary-vetted sample. The elevated occurrence inwards of 1 yr is explainable by increased stellar activity around these young hosts (see \paperII{}) rather than an inner population of BDs.

The detections are plotted at their published masses and orbital periods: 51~Eri~b \citep{elliott_2024, denis_2026}, HD~95086~b \citep{derosa_2016, chauvin_2018}, HR~8799~c, d and e \citep{zurlo_2022, brandt_2021b}, HD~984~B and PZ~Tel~B \citep{bernardi_2026}, and HR~2562~B \citep{konopacky_2016, zhang_2023}. Only HR~8799~e, HD~984~B and PZ~Tel~B have dynamical masses; the dynamical mass of PZ~Tel~B ($24^{+10}_{-7}$\,\mjup{}) is far below the 64\,\mjup{} adopted by \citet{nielsen_2019}, and the luminosity-based mass of HR~2562~B ($30 \pm 15$\,\mjup{}) exceeds its dynamical upper limit of 18.5\,\mjup{} \citep{zhang_2023}.

We compare in the period range where astrometry measures mass and separation separately, $P = 3$--94\,yr (the Letter's window; $a \approx 2.5$--26\,au at these hosts), and quote the GPIES rates at 10--100\,au, so the two overlap in separation only at 10--26\,au (red outline in Figure~\ref{fig:gpies}(a), shading in (b)). Beyond $\sim$100\,yr astrometry constrains mainly the acceleration $m/a^2$, which smears each star's posterior along that degeneracy and appears to fill in the brown dwarf desert.
For the 97 GPIES targets in our sample above 1.5\,\msun{}, our planet rate is $5.6^{+2.6}_{-1.9}\%$ (from our refit to all 238 GPIES targets). Over our full population, weighted by the GPIES host masses, the rate is $14.7^{+0.8}_{-0.9}\%$. The rate reported by GPIES falls between these two values ($8.9^{+5.0}_{-3.6}\%$).

Based on the survey mean completeness of \citet{nielsen_2019}, our results fit to the same 238-star subset predict $1.7^{+0.9}_{-0.6}$ planet detections in their planet box around those 97 stars (3 observed, HR~8799~c, d and e; $P(N \ge 3) = 0.24$) and $5.9^{+2.1}_{-1.6}$ brown dwarfs around the 238 stars (3 observed; $P(N \le 3) = 0.16$).

Beyond these summary rates, Figure~\ref{fig:gpies}(b)--(e) compares the distributions themselves, over 3--100\,au and above 5\,\mjup{}, where our sensitivity at these host masses is good ($P \geq 3$\,yr and $q \geq 10^{-3}$, as in Figure~\ref{fig:funnel}). In panel (b), the expected number of companions versus semi-major axis is compared with GPIES. The rates essentially meet near $\approx10$ au where both methods are sensitive. As discussed in this Letter, the ability of our constraints to measure the period dependence of low mass-ratio companions is very weak---constraints from Gaia DR4 may significantly change this picture.

In mass (Figure~\ref{fig:gpies}(c)) the desert is clear in this window: our population falls to a minimum before rising again. The rate of brown dwarfs in this sample is elevated in the desert compared to the \citet{nielsen_2019} power law, but we attribute much of this disagreement to the definition of brown dwarf adopted in the GPIES power laws (13--80\,\mjup{}): this window overlaps with the rising low mass-ratio companion rate around higher mass stars, smearing occurrence into this window. As argued in this Letter, the brown dwarf desert is better described as a mass-ratio desert, and this window straddles both sides for these host masses.

The host-mass dependence of planetary and brown-dwarf companions (by the GPIES definitions) is then plotted in Figure~\ref{fig:gpies}(d)--(e). Following \citet{nielsen_2019}, we fit occurrence $\propto (M_\star/1.75\,\msun)^\gamma$ to our host-mass strata. We find a very similar trend in increasing planet rate with host mass (ours $\gamma = 2.7 \pm 0.2$; GPIES $\gamma = 2.0 \pm 1.0$). For brown dwarfs, we find a similar neutral to slightly negative trend (ours $\gamma = -0.3 \pm 0.3$ for K--A hosts; GPIES $-0.9 \pm 1.5$), the notable aforementioned exception being elevated brown dwarf rates around B-type stars, attributed to the 13--80\,\mjup{} window picking up the rising low mass-ratio companion occurrence below the mass-ratio desert.

Future work could combine direct imaging constraints in the per-star posteriors (following \citet{Ruffio_2018} and e.g. \citet{octofitter} or \citet{llopsayson2026_epseri_nircam}) to perform joint astrometric and direct imaging demographics using this framework.

%% file: app-1.tex
\clearpage\section{Detailed Methodology}\label{ap:methods}

\begin{figure*}[t]
\centering
\includegraphics[width=\textwidth]{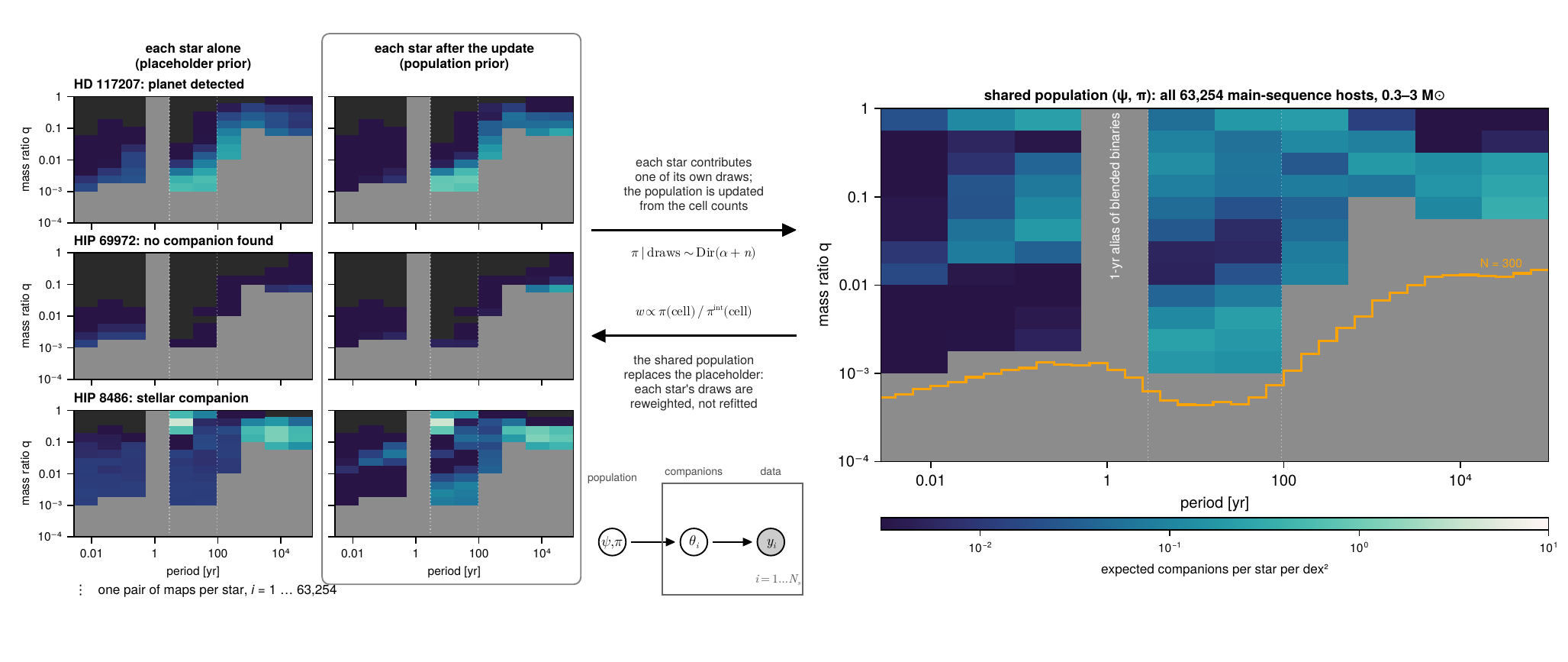}
\caption{A hierarchical population model of companions around our full 63,254 star sample. The heatmaps plot the posterior median on the expected number of companions per star, per dex in orbital period, per dex in companion mass-ratio. The leftmost column shows the independent posteriors of three illustrative stars from \paperII{}. Moving rightwards, the next column shows how these posteriors are updated by the joint population model. The pooled information across all stars updates the priors of individual systems to the shared population distribution and sharpens the inferred parameters of companions. The rightmost panel shows the inferred population-wide distribution of companions as a function of orbital period and companion mass-ratio, in units of companions / star / decade of period / decade of mass-ratio.}
\label{fig:update}
\end{figure*}

\subsection{Models}\label{ap:octopodes}
\subsubsection{System model}\label{ap:system-model}

Each star is first fit independently with the system model of \paperII{}: a trans-dimensional Octofitter \citep{octofitter} model with up to $\Nmax = 3$ companions, sampled with non-reversible parallel tempering \citep{pigeons}. Its parameters, priors and likelihood terms are listed in \paperII{} (its Section~2 and priors table). For the population model only the following properties of the system model matter:
(i) the number of active companions $N_s \in \{0, \dots, \Nmax\}$, $\Nmax = 3$, has the interim prior $\psiint = (0.500, 0.278, 0.144, 0.078)$;
(ii) given $N_s$, the companions' periods and mass ratios are independent and log-uniform, $P_k \sim \mathrm{LogU}(1\,\mathrm{d}, 10^5\,\mathrm{yr})$  and $q_k \sim \mathrm{LogU}(10^{-5}, 1)$, so the interim probability that a companion falls in cell $\ell$, $\piint_\ell$, is the fraction of that log-uniform prior inside the cell;
(iii) the active companions are exchangeable.
We use the terminology \emph{system trace} to describe a collection of samples coming from one Octofitter run based on interim priors.
By default, the \emph{region of interest} is the support of the prior on $(P, q)$; it can be restricted when 
going from system traces to the population model (we restricted to $q \le 1$), but one has to ensure the effect of filtering is not biasing the prior on the 
number of companions. One simple method to achieve this is to take into account the effect of the filtering on the specification of the interim prior passed to the population model.

\subsubsection{Population model}\label{ap:model}

The population model replaces the interim priors $\psiint$ and $\piint$ of the system model with population-level distributions that are inferred jointly from all $S$ systems (Figure~\ref{fig:update}). See Table~\ref{tab:joint-notation} for a summary of the mathematical notation, and Figure~\ref{fig:joint-gm} for a directed graphical model of the joint model.   The parameter $\psi_n$ gives the population probability that a system hosts $N_s = n$ companions. The cell distribution $\pi$ gives the probability that a companion falls in cell $\ell$ of a grid over $(\log_{10} P, \log_{10} q)$. In this work the grid has $L = 200$ cells: 10 in period, tiling the full prior range, by 20 in mass ratio, of 0.25\,dex from $q=10^{-5}$. 

Given $\psi$ and $\pi$, systems are independent; given $N_s$, each companion's cell is an independent draw from $\pi$; and the data $y_s$ depend on the population only through the companions of system $s$.

\begin{deluxetable}{lp{12.5cm}}
\tablecaption{Notation used in the population model.\label{tab:joint-notation}}
\tablehead{\colhead{Symbol} & \colhead{Description}}
\startdata
$s \in \{1, \dots, S\}$ & System index \\
$k \in \{1, \dots, \Nmax\}$ & Companion index (companion slots $b, c, d$ of \paperII{}) \\
$\ell \in \{1, \dots, L\}$ & $(P, q)$ cell index \\
$y_s$ & Data observed for system $s$ \\
$N_s \in \{0, 1, \dots, \Nmax\}$ & Number of companions in the $(P,q)$ region of interest for system $s$ \\
$x_{s,k} \in \{1, \dots, L\}$ & Cell membership of companion $k$ of system $s$ \\
$x_s = (x_{s,1}, \dots, x_{s,N_s})$ & Vector of all companions' cell memberships for system $s$ (or an empty vector if $N_s = 0$) \\
$\psi = (\psi_0, \psi_1, \dots, \psi_{\Nmax})$ & Companion count distribution ($\sum_n \psi_n = 1$, $\psi_n \ge 0$) \\
$\pi = (\pi_1, \pi_2, \dots, \pi_{L})$ & Cell membership distribution ($\sum_\ell \pi_\ell = 1$, $\pi_\ell \ge 0$) \\
$\psiint$, $\piint$ & Interim priors on $N_s$ and $x_{s,k}$ used by the system model \\
$\alpha = (\alpha_1, \dots, \alpha_L)$ & Dirichlet concentration of the prior on $\pi$ \\
\enddata
\end{deluxetable}
\onecolumngrid 

\begin{figure}[t]
	\centering
	\begin{minipage}[c]{0.35\textwidth}
	\centering
	\begin{tikzpicture}[
		latent/.style={circle, draw=black!80, thick, minimum size=1.0cm},
		observed/.style={circle, draw=black!80, fill=gray!20, thick, minimum size=1.0cm},
		arrow/.style={-Stealth, thick, shorten >=2pt, shorten <=2pt},
		plate/.style={draw=black!60, rectangle, thick}
		]
		\node[latent] (psi) at (2,4) {$\psi$};
		\node[latent] (eta) at (4,4) {$\pi$};
		\node[latent] (ns) at (2,2) {$N_s$};
		\node[latent] (xs) at (4,2) {$x_{s}$};
		\node[observed] (ys) at (4,0) {$y_s$};
		\draw[arrow] (psi) -- (ns);
		\draw[arrow] (eta) -- (xs);
		\draw[arrow] (xs) -- (ys);
		\draw[arrow] (ns) -- (xs);
		\node[plate, fit=(ns) (ys) (xs), inner sep=12pt] (plates) {};
		\node[anchor=south east, font=\small] at (plates.south east) {$s$};
	\end{tikzpicture}
	\end{minipage}
	\begin{minipage}[c]{0.3\textwidth}
	\begin{align*}
		\psi &\sim \text{Dirichlet}(\underbrace{1, 1, \dots, 1}_\text{length $\Nmax + 1$}) \\
		\pi &\sim \text{Dirichlet}(\alpha) \\
		N_s | \psi &\sim \text{Categorical}_{0:\Nmax}(\psi) \\
		x_s  &= (x_{s, 1}, \dots, x_{s, N_s}) \\
		x_{s,k} | \pi &\sim \text{Categorical}_{1:L}(\pi) \\
		y_s | x_s &\sim p(\cdot | x_s)
	\end{align*}
	\end{minipage}
	\caption{Directed graphical model for the joint population model.}
	\label{fig:joint-gm}
\end{figure}

\subsubsection{Priors on the population}\label{ap:priors}

The prior on $\psi$ is always uniform (i.e., a Dirichlet with each concentration hyper-parameter set to one). For $\pi$ we use two priors. The flat Dirichlet, $\alpha_\ell = 1$, is used for the brown dwarf desert map, where each host-mass stratum is fit on its own. For occurrence rates as a function of metallicity, we instead use informative priors, fit using the following two step process inspired by a two level hierarchical model. Suppose we are interested in a given subset $G$ of the systems (defined by metallicity here). The idea is to use all the systems excluding $G$ to first build an informative prior. More precisely, the first step is a flat-prior fit to the whole pool with $G$ removed; its posterior mean $\pi^\mathrm{ref}$ summarises how stars in general distribute their companions. Next, we use the posterior of the first step to construct the prior for $\pi$ in the second step, as a Dirichlet with updated parameters $\alpha$. 

\subsection{Approximating the posterior distribution}\label{ap:mcmc}

\subsubsection{Joint MCMC sampler}\label{sec:imh-sampler}

Assume we have the system traces of all $S$ systems. Here we describe how these can be combined into an MCMC sampler for the joint model. The method is closely related to the sampler described in \citet{lunn_2013}.  

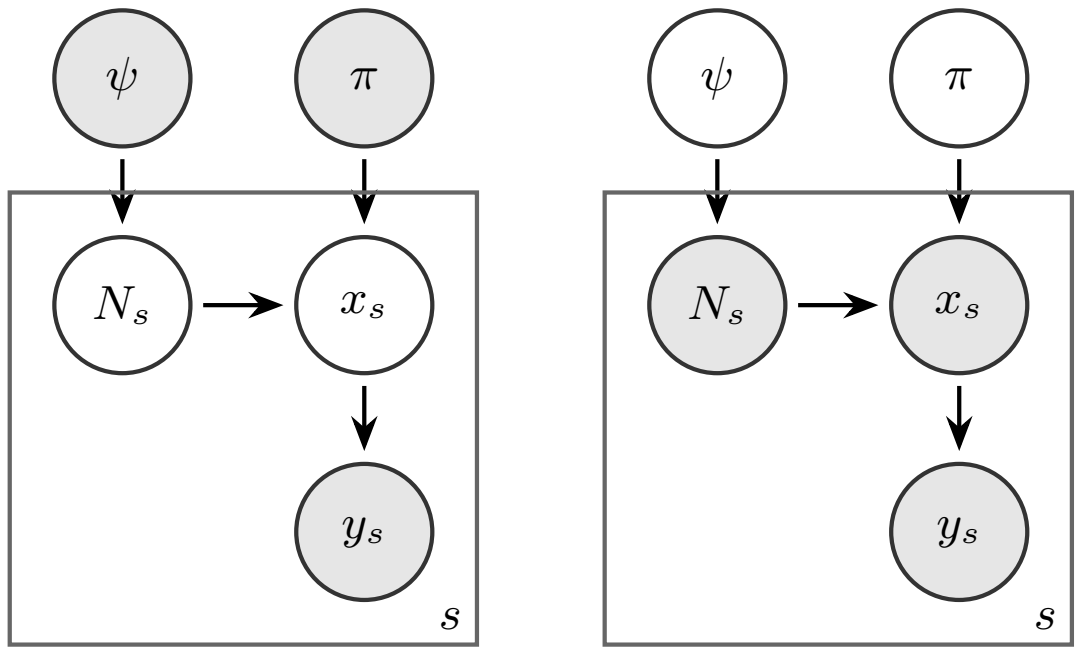
\begin{figure}[t]
\centering
\begin{tikzpicture}[
	latent/.style={circle, draw=black!80, thick, minimum size=0.9cm},
	observed/.style={circle, draw=black!80, fill=gray!20, thick, minimum size=0.9cm},
	arrow/.style={-Stealth, thick, shorten >=2pt, shorten <=2pt},
	plate/.style={draw=black!60, rectangle, thick}
	]
	\node[observed] (psi) at (0,3) {$\psi$};
	\node[observed] (pi) at (1.6,3) {$\pi$};
	\node[latent] (ns) at (0,1.5) {$N_s$};
	\node[latent] (xs) at (1.6,1.5) {$x_{s}$};
	\node[observed] (ys) at (1.6,0) {$y_s$};
	\draw[arrow] (psi) -- (ns);
	\draw[arrow] (pi) -- (xs);
	\draw[arrow] (xs) -- (ys);
	\draw[arrow] (ns) -- (xs);
	\node[plate, fit=(ns) (ys) (xs), inner sep=8pt] (plates) {};
	\node[anchor=south east, font=\small] at (plates.south east) {$s$};
\end{tikzpicture}
\hspace{0.6cm}
\begin{tikzpicture}[
	latent/.style={circle, draw=black!80, thick, minimum size=0.9cm},
	observed/.style={circle, draw=black!80, fill=gray!20, thick, minimum size=0.9cm},
	arrow/.style={-Stealth, thick, shorten >=2pt, shorten <=2pt},
	plate/.style={draw=black!60, rectangle, thick}
	]
	\node[latent] (psi) at (0,3) {$\psi$};
	\node[latent] (pi) at (1.6,3) {$\pi$};
	\node[observed] (ns) at (0,1.5) {$N_s$};
	\node[observed] (xs) at (1.6,1.5) {$x_{s}$};
	\node[observed] (ys) at (1.6,0) {$y_s$};
	\draw[arrow] (psi) -- (ns);
	\draw[arrow] (pi) -- (xs);
	\draw[arrow] (xs) -- (ys);
	\draw[arrow] (ns) -- (xs);
	\node[plate, fit=(ns) (ys) (xs), inner sep=8pt] (plates) {};
	\node[anchor=south east, font=\small] at (plates.south east) {$s$};
\end{tikzpicture}
\caption{Structure of the two within-Gibbs updates used to sample from the joint model's posterior distribution. Left:  independence-within Gibbs moves informed by first running one MCMC for each system. Right: conjugate updates of the population parameters given the rest.}
\label{fig:gibbs}
\end{figure}

The joint MCMC algorithm alternates between two within-Gibbs samplers (Figure~\ref{fig:gibbs}): (1) updating $N_s, x_s$ given the rest, and (2), updating $\psi, \pi$ given the rest; Figure~\ref{fig:update} shows their effect on three illustrative stars.

Update (2) is a straightforward conjugate update:
\begin{align*}
	\psi | \text{rest} &\sim \text{Dirichlet}\Big( 1 + \sum_s \mathbf{1}[N_s = 0], \dots, 1 + \sum_s \mathbf{1}[N_s = \Nmax]  \Big) \\
	\pi | \text{rest} &\sim \text{Dirichlet}\Big( \alpha_1 + \sum_{s,k} \mathbf{1}[x_{s,k} = 1], \dots, \\
	&\qquad\qquad\quad \alpha_L + \sum_{s,k} \mathbf{1}[x_{s,k} = L]  \Big),
\end{align*}
where the sums over $k$ run over the $N_s$ active companions of each system.

After the count statistics have been gathered, the running time of this update does not grow with $S$, making this step very fast.

For update (1), we first use conditional independence to break down the problem into $S$ independent updates. For each system $s$, consider an ``independence samplers within Gibbs'' algorithm where the proposal comes from the system traces.
The system trace of system $s$ targets the following distribution:
\begin{align*}
	q_s(N_s, x_s) = p(N_s, x_s | y_s, \psiint, \piint),
\end{align*}
where $\psiint$ and $\piint$ are the interim priors of the system model. Therefore the acceptance ratio for the independence sampler works as follows: propose $N'_s, x'_s \sim q_s$, and accept if a uniform $[0, 1]$ random number is lower than
\begin{align*}
	&\frac{p(\psi, \pi, N'_s, x'_s, y_s)}{p(\psi, \pi, N_s, x_s, y_s)} \frac{q_s(N_s, x_s)}{q_s(N'_s, x'_s)} \\
	&= \frac{p(N'_s | \psi) p(x'_s | N'_s, \pi) p(y_s | x'_s)}{p(N_s | \psi) p(x_s | N_s, \pi) p(y_s | x_s)} \frac{p(N_s | \psiint) p(x_s | N_s, \piint) p(y_s | x_s)}{p(N'_s | \psiint) p(x'_s | N'_s, \piint) p(y_s | x'_s)} \\
	&= \frac{p(N'_s | \psi) p(x'_s | N'_s, \pi) }{p(N_s | \psi) p(x_s | N_s, \pi)} \frac{p(N_s | \psiint) p(x_s | N_s, \piint) }{p(N'_s | \psiint) p(x'_s | N'_s, \piint) } \\
	&= \frac{r(N'_s) r(x'_s)}{r(N_s) r(x_s)},
\end{align*}
where $r(n) = \psi_n / \psiint_n$ and $r(x) = \prod_{k = 1}^{|x|} \pi_{x_k} / \piint_{x_k}$.
Here $\piint_\ell$ is the interim prior mass of cell $\ell$. 
In practice, we approximate the step where we sample from $q_s$ by drawing one of the system trace iterations uniformly at random, with replacement, independently for every system and every iteration. 

The output of the sampling algorithm is a set of MCMC samples.
We call the collection composed of these output samples, the joint trace, to differentiate it from the system traces.
Let $i \in \{1, 2, \dots, I\}$ denote the iteration index of the joint MCMC
algorithm. The output of the joint MCMC algorithm consists of matrices of real-valued sampled parameters, $\psi_n\mciter, \pi_\ell\mciter$, and
matrices of system states $N_s\mciter, x_s\mciter$.

As with any MCMC algorithm, these samples can be used to approximate posterior expectations via Monte Carlo averages: for
any integrable test function $f$ of interest,
\begin{align}\label{eq:tests}
	\frac{1}{I} \sum_{i=1}^I f\left(\psi\mciter, \pi\mciter, N\mciter, x\mciter\right) \approx \E[f(\psi, \pi, N, x) | y],
\end{align}
where $N = (N_1, \dots, N_S)$, $x = (x_1, \dots, x_S)$.
In the next section, we describe several examples of sample post-processing
methods based on different choices of test functions $f$.

\subsubsection{Post-processing of the joint sampler output}

By default, all post-processing discards the first 50\% of the samples to reduce
the bias of the Monte Carlo estimators (a typical strategy in the MCMC
literature known as warm up or burn in).

Most post-processing operations are obtained by straightforward choices of
test function $f$ described in Section~\ref{sec:imh-sampler}.  For example,
consider estimation of the posterior probability that system $s$ has exactly one companion in the
region of interest, $p(N_s = 1 | y)$.
To estimate $p(N_s = 1 | y)$,
we use:
\begin{align*}
	f_{s, 1}\left(\psi, \pi, N, x\right) = \mathbf{1}[N_s = 1],
\end{align*}
and plug-in $f = f_{s, 1}$ into Equation~(\ref{eq:tests}). 

The occurrence rates of the main text use the test function $f_W(\psi, \pi) = \E_\psi[N] \sum_{\ell \in W} \pi_\ell$, with $\E_\psi[N] = \sum_n n\,\psi_n$: the expected number of companions per star in a window $W$ of cells. 

%% file: app-2.tex
\clearpage
\section{Robustness Checks}\label{ap:robust}

Here we perform checks to see if the inferred population demographics are impacted by sources of noise or miscalibration that should not \emph{a priori} have significant impacts on inferred companion rates.

\subsection{Color: Main Sequence vs Red-Giant Branch}\label{ap:ms-rgb}

\begin{figure*}
\centering
\includegraphics[width=\textwidth]{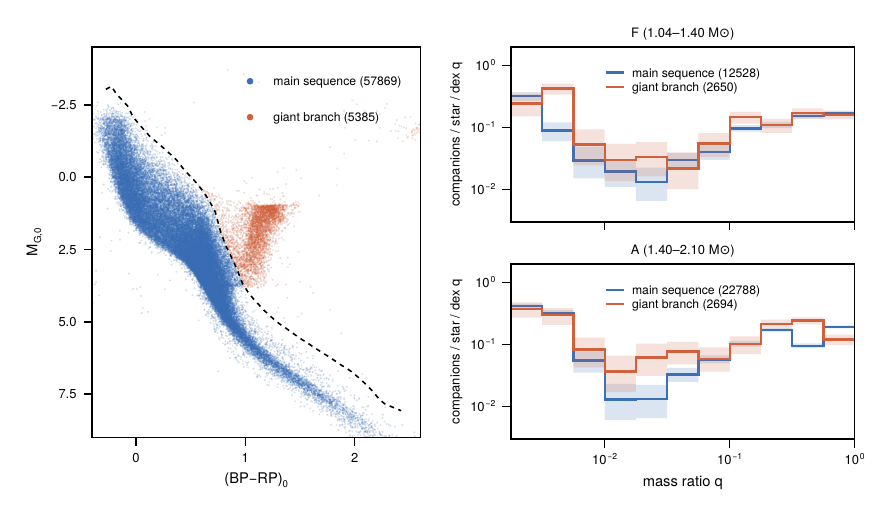}
\caption{Calibration check on color and evolutionary state. \textbf{Left:} extinction-corrected color--magnitude diagram of the 63,254-star sample, split into main-sequence (blue; 57,869) and giant-branch (orange; 5,385) hosts at 1.5\,mag above the bright edge of the MIST main sequence (dashed). \textbf{Right:} companion rates per star per dex of mass ratio, summed over 3--94\,yr, for the F (1.04--1.40\,\msun{}) and A (1.40--2.10\,\msun{}) host-mass strata, fit separately for each branch; steps are posterior medians and bands 16--84\,\%. Mass ratios start at the first bin wholly above the full stratum's 300-star compatibility limit, as masked in Figure~\ref{fig:desertmap}. Hosts of very different color give the same planet occurrence.}
\label{fig:ms-rgb}
\end{figure*}

We compare the BD desert result from the F and A host mass bins between the main sequence and giant branches in Figure \ref{fig:ms-rgb}. Because astrometric uncertainties in Gaia and Hipparcos are sensitive to color, a color-dependent mis-calibration or mis-specification in \paperI{} or \paperII{} could appear as an elevated population of spurious very low mass-ratio companions (near the noise floor) around either the red (giant branch) or blue (main sequence) populations. We find no such elevated low mass-ratio companion rate and few substantial differences between main sequence and giant branch companion rates. The exception we note is a higher floor to the BD desert around A-type giants, which may motivate future study of companion rates and migration as a function of evolutionary state.

\subsection{Scan Law: Galactic Latitude and Longitude}\label{ap:scanlaw}

\begin{figure*}
\centering
\includegraphics[width=\textwidth]{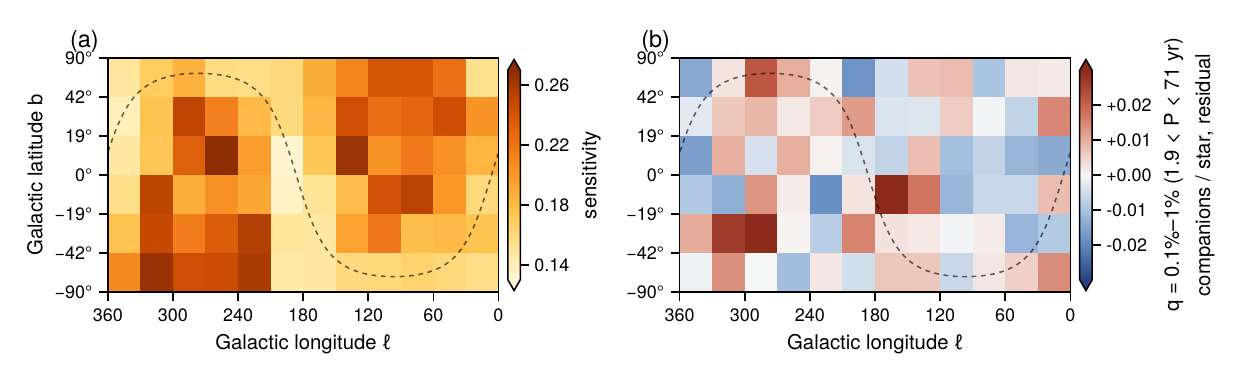}
\caption{Calibration check on sky position: the sensitivity to and deviation in the low mass-ratio companion rate as a function of galactic coordinate, in 72 equal-area cells (vertical axis uniform in $\sin b$). The sensitivity differs across the sky as expected, and no spurious low mass-ratio companion signal is detected as a function of position. (a) Each cell's sensitivity: the mean fraction of stars where $10^{-3} < q < 10^{-2}$ is detectable, tracing out the astrometric sensitivity of Gaia and Hipparcos. (b) Companions per star at $10^{-3} < q < 10^{-2}$ and $1.9 < P < 71$\,yr, each cell an independent population fit, minus a weighted linear trend with sensitivity (slope consistent with zero). No cell departs from the trend by more than $2.1\sigma$ despite 70 comparisons. The ecliptic is indicated by the dashed line.}
\label{fig:scanlaw}
\end{figure*}

In Figure \ref{fig:scanlaw}, we compare the overall companion rate across sky coordinates. Significant biases relating to the number of Gaia or Hipparcos observations or the orientation of those observations (both sky-position dependent) could appear as a spurious anisotropic low mass-ratio companion rate. Bins in sky position (and not e.g. selected on stellar kinematics) over $\sim1\,\rm{pc}-1\,\rm{kpc}$ are essentially homogeneous and our sample is sufficiently nearby that we do not probe any galactic scale variations in stellar populations. Indeed, we see no evidence of such anisotropy that would point to any gross scanlaw dependent noise misspecification.

